\documentclass[fleqn,10pt]{wlscirep}
\usepackage[utf8]{inputenc}
\usepackage[T1]{fontenc}
\usepackage[export]{adjustbox}
\usepackage[justification=justified]{caption}

\title{Spin distillation cooling of ultracold Bose gases}

\author[1,*]{Tomasz \'Swis\l{}ocki}
\author[2]{Mariusz Gajda}
\author[3]{Miros\l{}aw Brewczyk}
\author[2]{Piotr Deuar}

\affil[1]{Institute of Information Technology, Warsaw University of Life Sciences -- SGGW, ul. Nowoursynowska 159, PL-02786 Warsaw, Poland}
\affil[2]{Institute of Physics, Polish Academy of Sciences, Aleja Lotnik\'ow 32/46, 02-668 Warsaw, Poland}
\affil[3]{Wydzia\l{} Fizyki, Uniwersytet w Bia\l{}ymstoku, ul. K. Cio\l{}kowskiego 1L, PL-15245 Bia\l{}ystok, Poland}
\affil[*]{tomasz\_swislocki@sggw.edu.pl}

\begin{abstract}
We study the spin distillation of spinor gases of bosonic atoms and find 
two different mechanisms 
in ${}^{52}$Cr and $^{23}$Na atoms, both of which can cool effectively. 
The first mechanism involves dipolar scattering into initially unoccupied spin states and cools only above a threshold magnetic field. 
The second proceeds via equilibrium relaxation of the thermal cloud into empty spin states, reducing its proportion in the initial component. It cools only
 below a threshold magnetic field.
The technique was initially demonstrated experimentally for a chromium dipolar gas [B. Naylor {\it et al.}, Phys. Rev. Lett. {\bf 115}, 243002 (2015)], whereas 
here we develop the concept further and provide an in-depth understanding of the required physics and limitations involved. 
Through numerical simulations, we reveal the mechanisms involved and demonstrate that 
the spin distillation cycle can be repeated several times, each time resulting in a significant additional reduction of the thermal atom fraction.
Threshold values of magnetic field and predictions for the achievable temperature are also identified. 
\end{abstract}
\begin{document}

\flushbottom
\maketitle

\section{Introduction}

Since the time when the phenomenon of Bose-Einstein condensation was first predicted \cite{Bose,Einstein}, different methods of cooling have been developed, which eventually enabled the observation of the transition to a condensate. Laser cooling techniques, for example, allowed reaching microKelvin temperatures in atomic vapors \cite{laser_cooling}. With the application of evaporative cooling, the first condensates 
were obtained in a dilute atomic gas confined in a magnetic trap \cite{Anderson95,Davis95,Bradley97}. 
Due to the extremely low densities of the gas, its temperature had to be decreased well into the nanoKelvin range, to reach the phase-space density required for the transition. The lowest temperatures ever reached with Bose-Einstein condensates have been of the order of hundreds of picoKelvin \cite{Leanhardt03,Medley11}. 
Nonetheless, the usefulness of evaporative cooling tends to be limited to temperatures of the order of the chemical potential of the condensate $\mu\approx gn_0$, because evacuating atoms at energies of $\mu$ or less reduces the condensate at much the same rate as the thermal cloud.

The creation of multicomponent Bose-Einstein condensates \cite{Stamper-Kurn, Chapman01} has opened possibilities for reaching extremely low temperatures using different mechanisms that were not available in a single-component gas \cite{Stamper-Kurn,Erhard04,Medley11,Volchkov14,Hamilton14,Laburthe1,Ruhrig15}.  Naylor {\sl et al.} \cite{Laburthe1} have considered and demonstrated one of these possibilities. They propose to use the spin degrees of freedom to efficiently remove entropy in partially Bose-condensed spinor gases, aiming to reach temperatures below the current limitations set by standard evaporative cooling. In the case of ${}^{52}$Cr, the method involves preparing a polarized condensate in the lowest Zeeman state $m_s = -3$ via a sufficiently high applied magnetic field \cite{Perrin, Chu, Greif}.  
Then, the magnetic field is rapidly reduced to a level that allows a depolarization process to occur\cite{Laburthe1, Hart, Salomon} mediated by dipolar interactions. If the Zeeman energy is of the same order as the thermal kinetic energy of the system ($k_B T \gtrsim|g_L | \mu_B |\mathbf{B}|$), thermal population of higher Zeeman states becomes possible
\footnote{%
Here, $k_B$ is the Boltzmann constant, $T$ the temperature, \mbox{$g_L\approx-2$} the Land\'e g factor, $\mu_B$ the Bohr magneton and $\mathbf{B}$ the external magnetic field.}.
Thermal atoms are much more likely to depolarize than condensate atoms because of the threshold energy for this process set by the Zeeman energy. Therefore, the condensate fraction in the lowest $m_s=-3$ state grows. 
Once higher spinor components have been populated thus, they are released from the trap 
using for example, magnetic field gradient techniques.  
Such a spin filtering method leads then to a decrease of the thermal fraction and entropy of the remaining atoms in $m_s = -3$.
The experiment demonstrated cooling of a ${}^{52}$Cr gas after one such cycle when starting in a temperature range $(0.5-0.85)T_c$.
It was conjectured that iteration of such cycles, with magnetic field adjustment after each step, could allow one to greatly reduce the temperature. 
The low temperature limiting factors are not very obvious, however.
It was also suggested in \cite{Laburthe1} via a 
simplified calculation that the \emph{quadratic} Zeeman effect could be used for the cooling of $^{23}$Na, without the need for dipolar interactions. 

Here we report on advanced numerical simulations made using classical field methods at realistic temperatures and experimental parameters, which robustly confirm the above conjectures. They also help us explain in some detail the physical underpinnings and conditions under which the spin distillation cooling can take place. We find that the physics of the cooling differs between chromium and sodium gases, as do the fundamental limitations on the processes in question. 

The paper is organized as follows: We first introduce the model in Sec.~\ref{Model}, including specific aspects of the dipolar ${}^{52}$Cr and contact interacting $^{23}$Na systems. Results are then presented in Sections~\ref{Chromiumsection} and~\ref{Sodiumsection} for chromium and sodium, respectively. Each demonstrates one of the two mechanisms we have identified. General remarks and a summary are given in Sec.~\ref{Conclusions}.

\section{Model}\label{Model}

We consider two systems: $^{52}$Cr in the $s = 3$ state and $^{23}$Na in the $F = 1$ hyperfine state. 
The gases are at temperatures that are a sizable fraction of the critical temperature $T_c$, so they are modeled by a complex classical field, which is the most tractable and accurate model under these conditions. Several complementary techniques have been developed \cite{Davis,Goral,Bradley,Connaughton,Stoof99,Berloff,Witkowska,Gawryluk17,DP19}, reviewed in \cite{Brewczyk07,Blakie08,Proukakis08}.

\subsection{Initial thermal states}
The cooling simulations start with thermal clouds of atoms in a 3d harmonic trap  polarized in a single
spin state, which is $m_s=-3$ for for chromium and  $m_F=0$ for sodium. 
Therefore a single-component stochastic Gross-Pitaevskii equation (SGPE) \cite{Gardiner03,Stoof99} 
 provides a convenient route to directly obtain thermal ensembles at a chosen temperature and particle number $N$ \cite{Proukakis08,Swislocki16}. 
One evolves the complex field $\Psi({\bf r})$ in technical time $\tau$ according to the equation
\begin{eqnarray} \label{GPE}
i\hbar \frac{\partial \Psi ({\bf r},\tau)}{\partial \tau} =  \left(1 - i\gamma \right) \Big[-\frac{\hbar^2}{2m} \nabla^2 + \frac{1}{2} m\, \sum_j\omega_j^2 r_j^2 + \mathcal{H}_I(\Psi) - \mu \Big] \Psi({\bf r}, \tau) + \sqrt{2 \hbar \gamma k_B T}\, \eta({\bf r},\tau) ,
\end{eqnarray}
until a stable fluctuation of observables in time around a mean is obtained --- i.e. the stationary ensemble.		
Eqn. (\ref{GPE}) is the classical field equation for a field in thermal and diffusive contact with a reservoir at temperature $T$ and chemical potential $\mu$. 
Here $m$ is the atomic mass, $\omega_j$ the trap frequencies in directions $j=\{x,y,z\}$ (with coordinates $r_j$), 
and 
$\gamma$ is a dimensionless coupling strength to the reservoir.
For our system the trap frequencies were $\omega_j/2\pi=(250,300,215)\,$Hz, in all cases.
The $\eta$ is a complex white noise field with mean zero and variances $\langle\eta^*({\bf r}, \tau)\eta({\bf r}', \tau')\rangle = \delta^{(3)}({\bf r} - {\bf r}')\delta(\tau - \tau')$, 
$\langle\eta({\bf r}, \tau)\eta({\bf r}', \tau')\rangle=0$.
Interaction energy is captured by the functional $\mathcal{H}_I(\Psi)$, which is 
\begin{eqnarray}
\mathcal{H}_I(\Psi) = g |\Psi({\bf r}, \tau)|^2 + H_d(\Psi)
\end{eqnarray}
with contact interactions in the 1st term, and dipolar interactions (only in chromium) in the second. 
For chromium, $H_d(\Psi) = m_s[H_d(\Psi({\bf r}))]_{11}$ with the latter given in Sec.~\ref{ddmel} 
and $g=g_6$ as described in Sec.~\ref{METH-CR};   
For sodium, $H_d=0$ and $g=c_0$ as  described in Sec.~\ref{METH-NA}.
The magnetic field and chromium dipoles are oriented  along the $z$ axis.

The standard SGPE approach models the quantum degenerate gas using a classical field that describes the low energy/high density part of the system on a basis set spanning modes with occupations $\gtrsim {\mathcal O}(1)$, and a reservoir that describes all remaining high energy modes with occupations $\lesssim {\mathcal O}(1)$. A common practice is to restrict the low energy subspace for $\Psi$ to all plane wave modes below a certain momentum cutoff $k_{\rm cut}$. We also follow this practice, setting the numerical lattice spacing $\Delta x=\pi/k_{\rm cut}$ using $\hbar k_{\rm cut}=0.78\sqrt{2\pi mk_BT}$ according to the optimal multi-observable choice in 3d from \cite{Pietraszewicz15}. 	
The box size was around $3.2R_{\rm TF}=(3.2/\omega_j)\sqrt{2\mu/m}$, 
 and was chosen to provide the experimentally measured condensate fraction. Thus, the mode space is chosen according to a balance of various observables. 
The simulations start from the neutral initial conditions of the vacuum state $\Psi({\bf r},0)=0$, and used $\gamma=0.5$,
an efficient value for finding equilibrium ensembles (they do not depend on $\gamma$ \cite{Gardiner03}).
$\mu$ is chosen to obtain the desired ensemble average particle number. 
Dependence on the two control parameters $T$ and $\mu$ strongly suggests that this approach should be equivalent to others based on the grand canonical statistical ensemble (as, for example, in \cite{Gawryluk17}).
Indeed, it can be demonstrated that the stochastic field fulfills the fluctuation-dissipation relation \cite{next work}
\begin{eqnarray} \label{kBT_test}
\langle N^2 \rangle -  \langle N \rangle^2 = k_B T \left ( \frac{\partial N}{\partial \mu} \right )_T  \,,
\end{eqnarray}
which is an indication of required thermal properties.
Differences between grand canonical and canonical ensembles are negligible in the interacting gas in the regime of interest \cite{Wilkens97,Kocharovsky16,Pietraszewicz17}.

Samples of the stabilized field $\Psi ({\bf r},\tau)$ in the stationary ensemble are used 
as the initial states for further evolution. 
This subsequent evolution, during which cooling cycles are carried out is best performed without the high energy bath, because of its inherently non-equilibrium nature and the long time scales involved. It proceeds then via a plain Gross-Pitaevskii equation (GPE) \cite{Brewczyk07,Blakie08}.

\subsection{Chromium gas in the $s = 3$ state}
\label{METH-CR}
The first system we focus on is a partially condensed gas of $^{52}$Cr atoms in the $s = 3$ state, as per the initial experiment \cite{Laburthe1}.
In the classical-field model the cloud is described by the seven-component spinor wave function 
$\boldsymbol{\psi}({\bf r}) = (\psi_3({\bf r}),\psi_2({\bf r}),\psi_1({\bf r}),\psi_0({\bf r}),\\ \psi_{-1}({\bf r}),\psi_{-2}({\bf r}),\psi_{-3}({\bf r}))^T$
with one component for each value of the spin projection $m_s\in\{-3,-2,\dots,2,3\}$ along the direction of the applied magnetic field ${\bf B}$.
It evolves according to the multicomponent Gross-Pitaevskii equation \cite{Swislocki14}:
\begin{eqnarray} \label{Crequ}
i\hbar \frac{\partial}{\partial t} \boldsymbol{\psi}({\bf r})  &=& \left( H_{\rm sp} + H_c + H_d \right)\boldsymbol{\psi}({\bf r}),
\end{eqnarray}
 where the energy functional consists of three parts. The first one represents the single-particle term 
\begin{eqnarray}
H_{\rm sp} = -\frac{\hbar^2}{2m} \nabla^2 + V_{\rm trap}({\bf r}) - \boldsymbol{\mu} \cdot {\bf B}
\end{eqnarray}
which includes the kinetic, potential $V_{\rm trap} = \tfrac{1}{2}m\sum_j\omega_jr_j$, as well as Zeeman energies, in which $\boldsymbol{\mu} = g_L\mu_B {\bf s} /\hbar$. 
Since the magnetic field direction is constant in the $z$ direction, $ - \boldsymbol{\mu} \cdot {\bf B} = -g_L\mu_Bm_s|{\bf B}|\approx 2\mu_Bm_s{\rm B}$.

The second term originates from the contact interactions. $H_c$ is a $7\times7$ matrix in spinor components. The matrix elements are evaluated in a two-atom basis consisting of $|s,m_1\rangle |s,m_2\rangle$ states, where $m_1,m_2\in\{-3,2,...,2,3\}$, and $|s,m_s\rangle$ is the state of a single atom with spin $s$ and projection $m_s$. The matrix elements are given by
\begin{eqnarray}
[H_{c}]_{m_1 m_1'} = \sum_{S} g_{\scriptscriptstyle S} \sum_{M=-S}^S  \sum_{m_2,m_2'}  \langle s,m_1;s,m_2|S M\rangle \langle SM|s,m_1';s,m_2'\rangle \, \psi^{*}_{m_2}({\bf r})  \, \psi_{m_2'}({\bf r})  \,.
\label{Hcab}
\end{eqnarray}
Here, the symbol $\langle s,m_1;s,m_2|S M\rangle$ is a Clebsch-Gordan coefficient and $|S M\rangle$ is the state of a pair of atoms with a total spin $S$ and spin projection $M$. Only $S=0,2,4,6$ channels are allowed,  $M=m_1+m_2=m_1'+m_2'$ and the interaction strengths characterizing the colliding atoms with a total spin $S$ are given as $g_{\scriptscriptstyle S}=4\pi\hbar^2a_S/m$. We take the scattering lengths to be: $a_0=91$, $a_2=-7$, $a_4=63$, and $a_6=102$ in units of the Bohr radius \cite{Pasquiou10,Kawaguchi06}.

The last term in Eq. (\ref{GPE}) 
describes the dipolar interactions. Since the spin projection of a pair of atoms colliding via dipolar forces can change at most by $2$ and the spin projection of a single atom changes maximally by $1$, the matrix $H_{d}$ becomes tridiagonal \cite{Swislocki14}. On the main diagonal one has $[H_{d}]_{m m}=m\, [H_{d}]_{11}$. The elements on the first diagonal below are $[H_{d}]_{m,m-1}=\sqrt{(4-m) (3+m)/12}\, [H_{d}]_{10}$. Sec.~\ref{ddmel} gives $[H_{d}]_{11}$ and $[H_{d}]_{10}$, 
and all other elements are determined by the requirement that $H_d$ be Hermitian.

Evolution starts with the equilibrated stochastic field $\Psi ({\bf r})$ obtained with Eq. (\ref{GPE}) placed in the $m_s=-3$ state as $\psi_{-3}({\bf r})$, and vacuum $\psi_{m_s}({\bf r})=0$ in the 
other spin states $m_s=-2,\dots,3$. 
Subsequent evolution follows 
Eq.~(\ref{Crequ}).

\subsection{Sodium gas in the $F = 1$ hyperfine state}
\label{METH-NA}

The second system we focus on is a thermal gas of ${}^{23}$Na atoms. Due to their low magnetic moment, their dipolar interactions will be not considered. This is a spin-1 system described by a spinor wave function 
with three components  $\boldsymbol{\psi}=(\psi_1,\psi_0,\psi_{-1})^T$ of spin projection $m_F=0,\pm 1$ along the externally applied magnetic field.

The model Hamiltonian of the system is simpler than in the case of ${}^{52}$Cr and 
contains no dipolar term. Therefore in the evolution equation (\ref{Crequ}), $H_d=0$. 
The contact part for the $j$th component, coming from (\ref{Hcab}) with $S=0,2$, can be written
\begin{equation} \label{Nac}
i\hbar\frac{\psi_j({\bf r})}{\partial t} = c_0 n({\bf r}) \psi_j({\bf r}) + c_2 {\bf F}\cdot \frac{\partial{\bf F}}{\partial\psi_j^*({\bf r})},
\end{equation}
in which 
$n({\bf r})=\sum_{m_F} n_{m_F}({\bf r})=\sum_{m_F} |\psi_{m_F}({\bf r})|^2$ is the local total atom density and ${\bf F}=(\psi^{\dagger}F_x\psi,\psi^{\dagger}F_y\psi,\psi^{\dagger}F_z\psi)$ is the local spin density.
The $F_{x,y,z}$ are the spin-$1$ matrices. The spin-independent and spin-dependent interaction 
coefficients are $c_0=4\pi \hbar^2 (2 a_2 + a_0)/3m$ and $c_2= 4\pi \hbar^2 (a_2 - a_0)/3m$, 
where $a_S$ is the $s-$wave scattering length for colliding atoms with total spin $S$ \cite{Isoshima99,Ketterle01}.
We take $a_0=50$ and $a_2=55$ in units of the Bohr radius $a_B$,  consistent with most determinations \cite{Crubellier99,Samuelis02,Knopp11}. 

Since  the Hamiltonian of an $F=1$ spinor gas is invariant  with  respect to a rotation  of  the  spin  vector  around  the  direction of the  magnetic field,  we consider only the quadratic Zeeman effect ~\cite{Matuszewski}.
The magnetic part of the Hamiltonian is
$
H_{\rm QZE} = -q \int d {\bf r} \ n_0({\bf r}),
$
where a constant offset has been dropped. $n_0$ is the atom density in the $m_F=0$ component, and 
$q = \alpha_q |{\bf B}|^2$ is the Zeeman energy. For sufficiently low magnetic fields,  $\alpha_q / 2\pi \hbar \approx 277\,$Hz/$G^2$ \cite{Gerbier}.
Then 
\begin{eqnarray}
H_{\rm sp} = -\frac{\hbar^2}{2m} \nabla^2 + V_{\rm trap}({\bf r}) - q\,\delta_{m_F,0}. 
\end{eqnarray}

Similarly to the chromium case, the state with the lowest effective magnetic energy (here $m_F=0$) is initially populated by the stochastic field $\Psi({\bf r})$ generated by (\ref{GPE}).
The other components start with a tiny seed of $\psi_{\pm1}({\bf r})=10^{-5}\Psi({\bf r})$.
The dynamics of the system is subsequently described by: 
\begin{eqnarray} \label{Naequ}
i\hbar\frac{\partial \psi_0}{\partial t}  &=& \left(-\frac{\hbar^2}{2 m} \nabla^2 + V_{\rm trap} + c_0\, n - q\, \right)\psi_0 + c_2\left[(n_1+n_{-1})\psi_0+2\psi_0^*\psi_1\psi_{-1}\right],\nonumber\\
i\hbar\frac{\partial \psi_{\pm 1}}{\partial t}  &=& \left(-\frac{\hbar^2}{2 m} \nabla^2 + V_{\rm trap} + c_0\, n \right)\psi_{\pm 1} + c_2\left[(n_{\pm 1}- n_{\mp 1} + n_0)\psi_{\pm 1}+\psi_{\mp 1}^*\psi_0^2\right].
\end{eqnarray}

\section{Chromium: condensate assisted dipolar scattering to higher spins}
\label{Chromiumsection}

We proceed as in the initial experiment \cite{Laburthe1} and begin with the chromium gas confined in the harmonic trap, with all atoms in 
the lowest Zeeman state of the system, $m_s = -3$. Dipolar collisions allow for a change of the magnetization of the system \cite{Ueda,Kawaguchi06,Gawryluk07,Gawryluk11,Swislocki11}. 
In fact, transfer of atoms to $m_s \ge -2$ components is possible only due to dipolar interactions \cite{Swislocki14}. 
Energies available in the cloud can surmount the Zeeman energy difference $|g_L\Delta m_s|\mu_B{\rm B}\sim2\mu_B{\rm B}$ when the magnetic field magnitude ${\rm B}=|{\bf B}|$ is sufficiently low, making possible
the population of higher Zeeman states. 
What one also wants for cooling is for thermal atoms to change spin state, but condensate atoms to remain in $m_s=-3$.
The conditions needed for this are twofold. First that thermal atoms can cross the Zeeman threshold, requiring
\begin{equation}\label{condCr}
2\mu_B{\rm B}\lesssim k_BT.
\end{equation}
Second that condensate atoms cannot, requiring that the ground state is ferromagnetic. The threshold magnetic field
above which the ground state of the system remains polarized in $m_s=-3$ \cite{Kechadi_thesis} is 
\begin{equation}
 {\rm B}\ \gtrsim\ B_{\rm th} = 
\frac{4\pi\hbar^2}{m \mu_B}\left[\frac{13}{66}\left(a_6-a_4\right)+\frac{5}{42}\left(a_4-a_2\right)\right]\, n,
\label{Bth}
\end{equation}
where $n$ is the atomic density. 
When the condensate peak density is used for $n$,
condition (\ref{Bth})
prevents all condensate atoms from changing spin state. 
We can safely omit dipolar contributions to condition (\ref{Bth})  because their input to condensate energy is minor \cite{Santos,Ho}.
The ratio of dipolar length ($a_{dd}=\mu^2 m/3\hbar^2$) to contact scattering length in the $m_s=-3$ state is $\epsilon_{dd}=a_{dd}/a_6=0.15 \ll 1$.

In each cooling cycle the initial cloud is allowed to evolve in-trap for 155ms. At the end of the cycle, all atoms in the higher $m_s\ge-2$ spin states are removed. Then a new cycle is begun starting with the remaining atoms in $m_s=-3$. Condensate fractions of the clouds are estimated by the spatial averaging technique \cite{Brewczyk07,Gawryluk08}. In this case, the cloud is averaged over the $z$ coordinate to construct an approximate one-body density matrix $\bar{\rho}^{(1)}_j(x,y,x',y',t) = \int dz\,\psi^*_j(x,y,z,t)\psi_j(x',y',z,t)$. The highest eigenvalue of $\bar{\rho}^{(1)}_j$ gives an estimate of condensate occupation in spin state $j$.
\begin{figure}[!bht]
\includegraphics[width=0.35\textwidth, angle=-90, center]{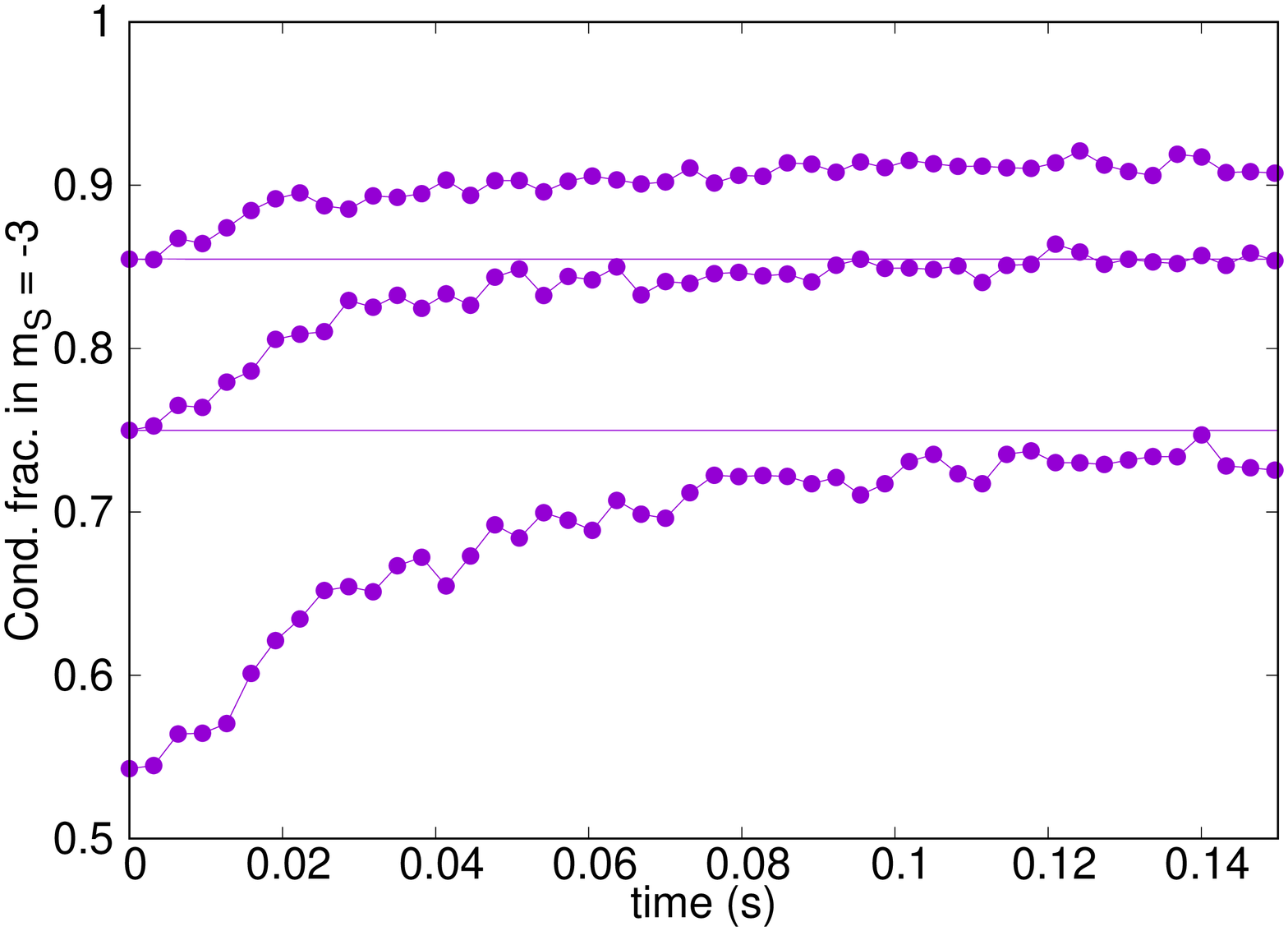}  \\ \vspace{0.5cm} \\
\includegraphics[width=0.47\textwidth, center]{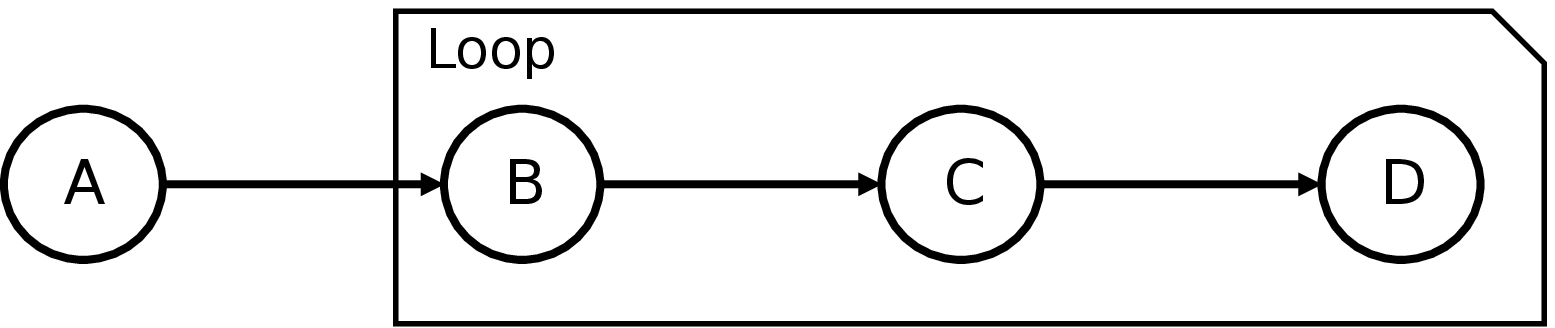}
\caption{             Upper frame:  Condensate fraction of ${}^{52}$Cr gas in the $m_s = -3$ Zeeman component as a function of time. Three sets of data represent three successive steps of the cooling procedure. Simulations were performed at magnetic fields chosen according to the rule: $2 \mu_B \mathrm{B} = k_B T_0$ using the initial temperature $T_0=235\,$nK at the beginning of the cooling procedure.  Hence, the magnetic field equals $1.75\,$mG for all cooling steps.              Lower frame: Schematic diagram of time sequence of cooling event: (A) -- a thermal sample of atoms is prepared, (B) -- active cooling begins, (C) -- active coolings stops, (D) -- unwanted atoms are removed. 
}
\label{fig:fig1}
\end{figure}

In Fig.~\ref{fig:fig1} three successive cycles of a spin distillation procedure of partially condensed ${}^{52}$Cr are presented. Initially the cloud contains $N\approx 20000$ atoms, with a condensate fraction of $0.55$, corresponding to a temperature of about $T_0=235\,$nK. These conditions are similar to the experiments \cite{Laburthe1}. In this simulation the magnetic field was kept at the same value ${\rm B}=k_BT_0/2\mu_B$ through all cycles, chosen as per the temperature condition in (\ref{condCr}) with the initial temperature. After three cooling steps the condensate fraction increases very strongly up to $0.9$. Notably, this simulation demonstrates that successive cooling cycles can give continued improvement. The condition (\ref{condCr}) eventually breaks down due to falling $T$. In fact no further cooling is seen after the third cycle, and with this protocol the final condensate fraction is 91\%, with $N = 13000$ atoms in total.

\begin{figure}[!h]
\includegraphics[width=0.35\textwidth, angle=-90, center]{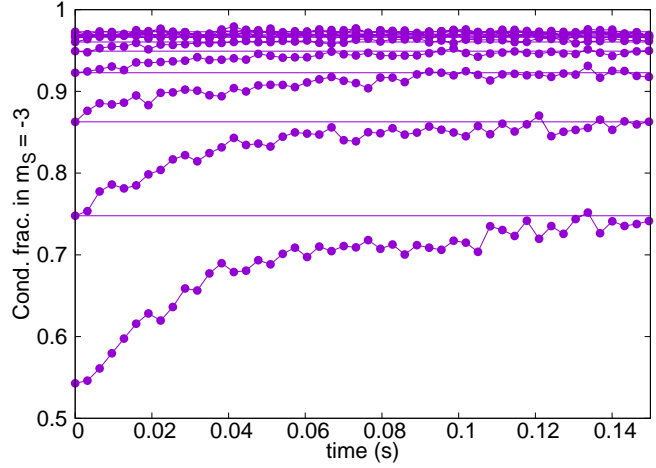}
\caption{Condensate fraction of ${}^{52}$Cr gas in $m_s = -3$, similarly to Fig.~\ref{fig:fig1}.
 Here, magnetic fields were chosen according to the rule: $2 \mu_B \mathrm{B} = \tfrac{1}{3}k_B T_{\rm cycle}$, with $T_{\rm cycle}$ determined by the value of condensed fraction at the beginning of a cycle.
The initial magnetic field was $0.58\,$mG.
}
\label{fig:fig100}
\end{figure}

It has been conjectured that much lower temperatures can be reached when the magnetic field is adapted to a running value of $T$ \cite{Laburthe1}. 
Fig.~\ref{fig:fig100} shows the progress when the magnetic field is adapted at the beginning of
each cooling cycle to the actual temperature value $T(t)$ according to 
the rule $2 \mu_B \mathrm{B} = \tfrac{1}{3} k_B T(t)$. The calculated condensate occupation $N_0(t)$ was used to obtain $T(t)=T^0_c(t) [1-N_0(t)/N(t)]^{1/3}$, where $T_c^0(N(t))$ is the the ideal gas critical temperature.  
Here, the final condensate fraction is about $0.97$ with $N = 12850$ atoms remaining.
\newline
\begin{figure}[!bht]
\includegraphics[width=0.35\textwidth, angle=-90, center]{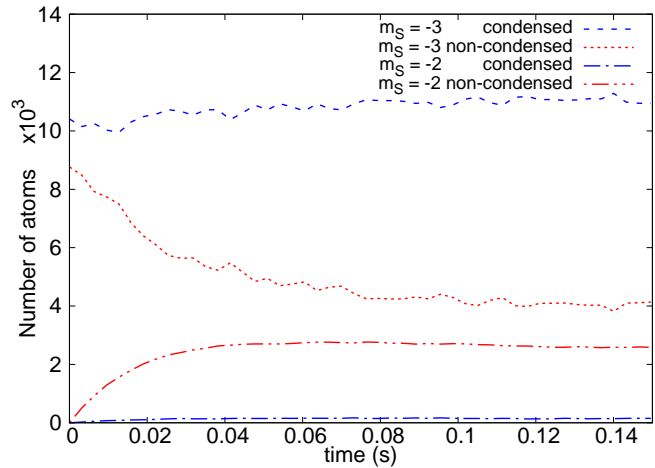}
\caption{Number of condensed (blue) and thermal (red) atoms in the $m_s = -3$ (dotted) and $m_s = -2$ (dash-dotted) components as a function of time, during the first cooling cycle depicted in Fig.~\ref{fig:fig1}.
}
\label{fig:Crfraction}
\end{figure}
\begin{figure}[!bht]
\centering
\includegraphics[width=0.30\textwidth, angle=-90]{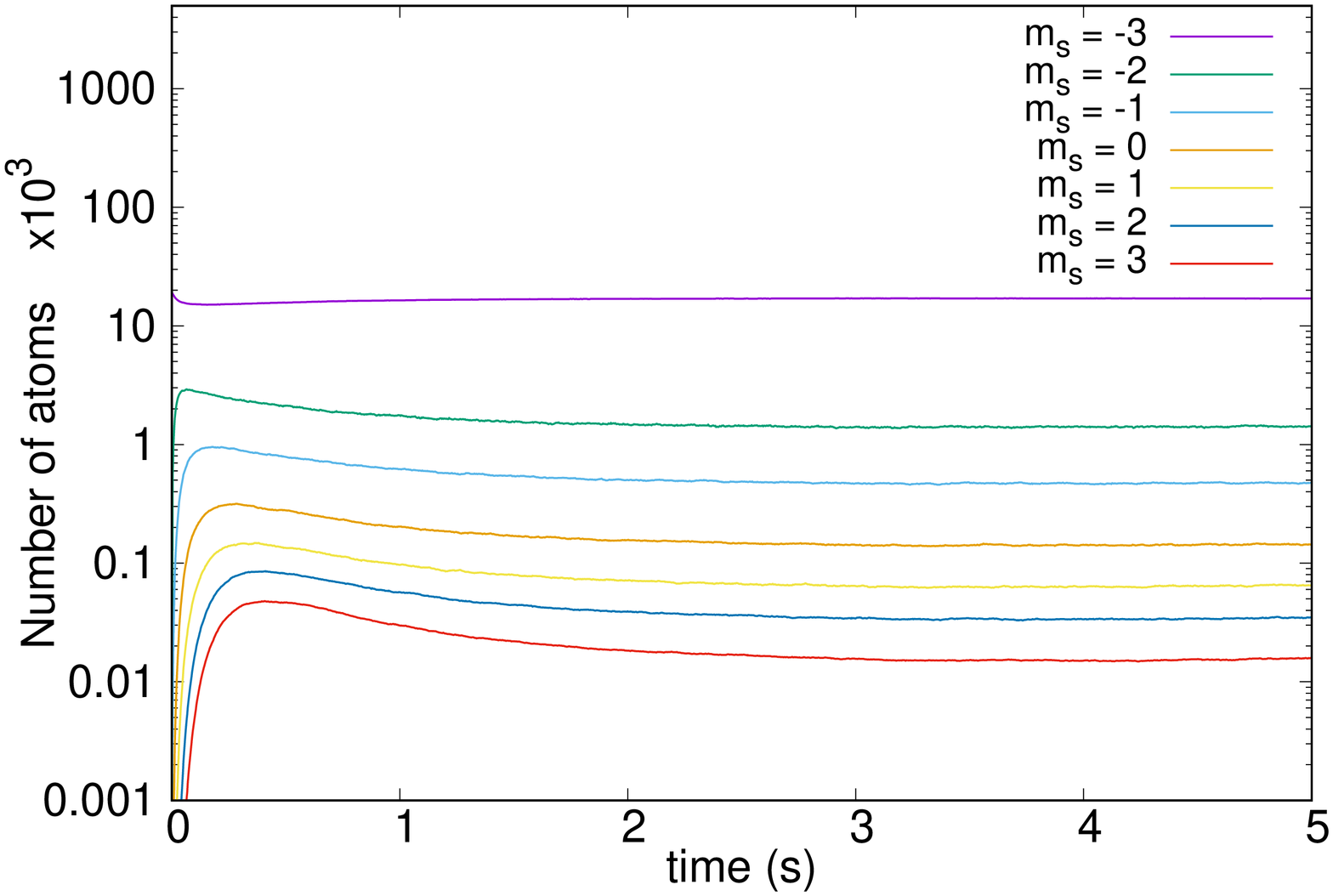}
\includegraphics[width=0.29\textwidth, angle=-90]{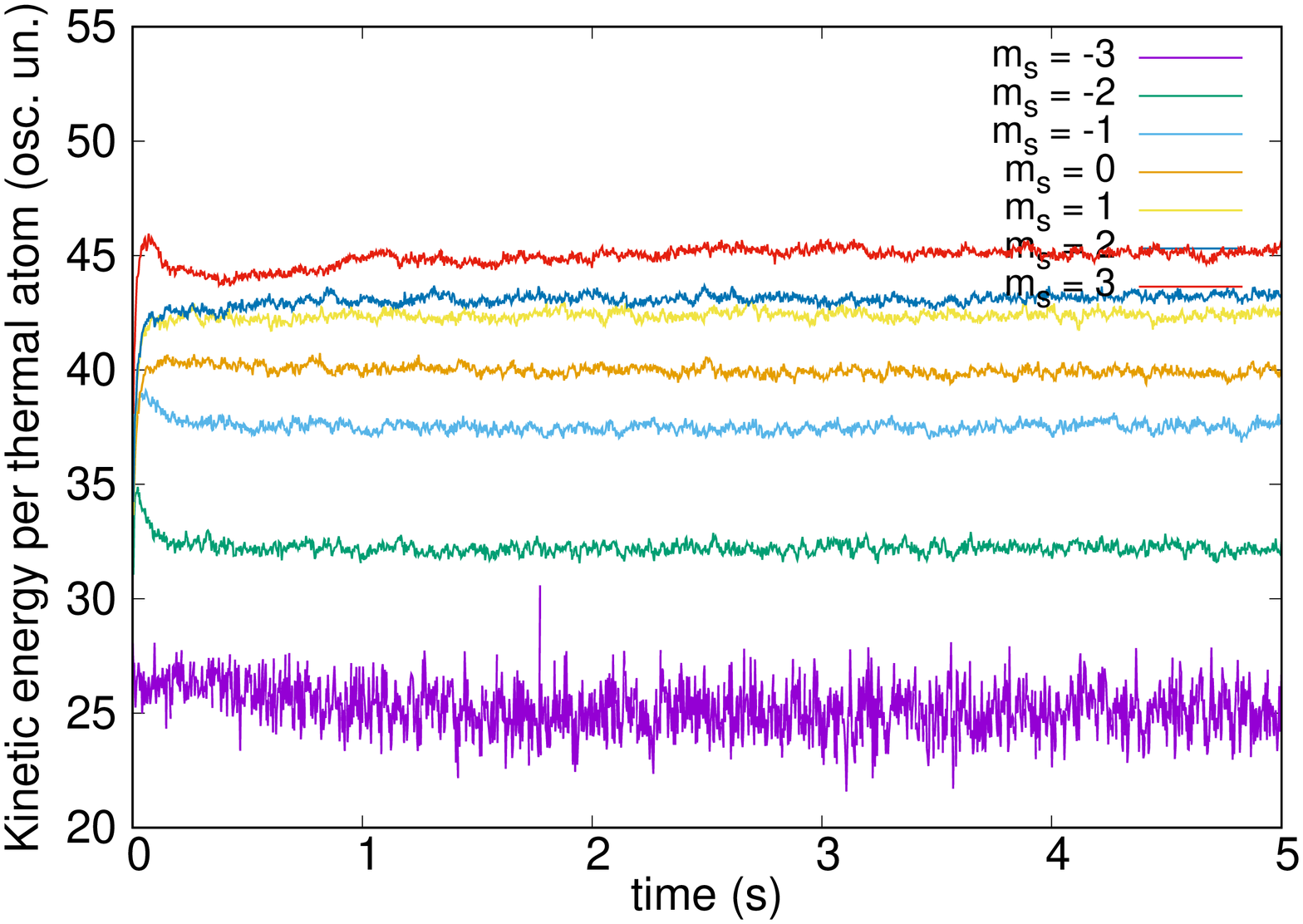}
\caption{
Populations (left) and kinetic energy per thermal atom (right, units of $\hbar\omega_x$) as a function of time, for all Zeeman components in the ${}^{52}$Cr simulations. A single long cooling step is made, with the same parameters as in Fig.~\ref{fig:fig1} (left) and as in Fig.~\ref{fig:fig100} (right).
}
\label{fig:Crkinetic}
\end{figure}

The dynamics is investigated in more detail in Figs. \ref{fig:Crfraction} and \ref{fig:Crkinetic} which plot the evolution of several quantities over a single cooling step. The first of these shows that the number of condensed atoms in the $m_s = -3$ state does not significantly change in time, while thermal cloud atoms migrate to higher $m_s$ components (primarily $m_s=-2$), which are purely thermal. This is as planned, and in agreement with the conditions (\ref{condCr})-(\ref{Bth}). Taking the Thomas-Fermi estimate of the peak density $n_{-3}^{\rm peak}$ in a fully condensed cloud \cite{Dalfovo99} as the maximum value of $n$,  we estimate that ${\rm B}_{\rm th}=0.29\,$ 
mG for our case.

An understanding of the cooling mechanism seen
can be summarized as follows:
\begin{enumerate}
\item 
Dipolar interactions populate the $m_s=-2$ state via the process $\psi^c_{-3}\,\&\,\psi^{\rm th}_{-3}\leftrightarrow\psi^c_{-3}\,\&\,\psi^{\rm th}_{-2}$, where the ${}^c$ and ${}^{\rm th}$ superscripts denote condensate and thermal fractions, respectively. Higher spin states are reached through $\psi^c_{-3}\,\&\,\psi^{\rm th}_{m_s}\leftrightarrow\psi^c_{-3}\,\&\,\psi^{\rm th}_{m_s+1}$, with $m_s=-2,-1,...,2$.
At ultracold temperatures Bose enhancement favors processes that involve the condensate, but all-condensate transitions like those seen in \cite{Swislocki14,Laburthe1} are ruled out by the Zeeman energy threshold.

\item 
No condensates appear in components $m_s=-2,-1,...,3$. The condensate remains only in the $m_s=-3$ state. This is because all but the lowest energy states are not populated highly enough and the saturation of the thermal gas is not attained. A crude estimation of the maximum number of atoms in excited states allowed by Bose statistics, based on the assumption of an ideal gas, gives about $9000$ atoms for our temperature $T=235\,$nK. Indeed, much less occupation of higher spin states is observed, shown in Fig.~\ref{fig:Crkinetic}, left panel.

\item 
Equilibration of thermal atoms within a spin component is seen: see Fig.~\ref{fig:Crkinetic}, right frame. The thermalization rates, due to contact interactions, are determined by the density $n$ of atomic cloud, the typical velocity $v$, and the cross section $\sigma$ through $\gamma = n \sigma v$. For the $m_s=-2$ Zeeman state, at the peak atomic density and at the critical temperature $\gamma\approx 40\,$s$^{-1}$ (\cite{Naylor}). Hence, a timescale of hundreds of milliseconds is needed to establish equilibrium within the $m_s=-2$ state. 
On the other hand, atoms in different spin states are not seen to equilibrate with each other, the kinetic energy per thermal atom does not equipartition between the Zeeman states:
Fig.~\ref{fig:Crkinetic} right frame. One reason is the low rate of spin changing collisions due to the small spatial overlap of thermal clouds appearing in different spin states. 
This occurs because higher spin states are low populated and the resulting clouds exhibit speckle-like density patterns due to thermal fluctuations.
Moreover, the equipartition is not guaranteed to hold in equilibrium for energy gaps of the order of $k_BT$ or more.

\item
Kinetic energy per thermal atom is higher in more energetic spin states. This can be due to the following 
 properties of dipolar interactions. For one, atoms transferred dipolarly to a higher spin state must carry nonzero angular momentum \cite{Swislocki10}. The higher the spin state, the more angular momentum needs to be carried. Therefore, the threshold energy consists of both Zeeman and rotational contributions. The energy associated with the rotation is next dissipated into the cloud increasing the kinetic energy present. The growth of the dipolar collision rate with velocity at low magnetic fields \cite{Pasquiou10,Hensler03} may also contribute to the highest energy atoms moving further up the spin ladder.
 Since spin-changing collisions are very low rate processes \cite{Naylor}, the net result is an increase of kinetic energy per thermal atom in successive spin states.

\end{enumerate}

Together, the conditions (\ref{condCr}) and (\ref{Bth}) set a limit for this cooling approach: $k_BT\gtrsim 2\mu_B{\rm B}_{\rm th}$. Substituting 
the peak density estimate
$n=\mu/g_6$, 
 we obtain 
\begin{equation}\label{limit1}
k_BT \gtrsim 
\left[\frac{13}{33}\left(\frac{a_6-a_4}{a_6}\right)+\frac{5}{21}\left(\frac{a_4-a_2}{a_6}\right)\right]
\mu.
\end{equation}
In particular, depending on the scattering lengths, 
this can be very far below $\mu$, which is the usual limiting value for standard evaporative cooling. For our present parameters, condition (\ref{limit1}) gives a very low value of 39nK.  

\section{Sodium: cooling thanks to redistribution of the thermal cloud}
\label{Sodiumsection}

The second case of a sodium gas in the $F = 1$ hyperfine state, was considered briefly by \cite{Laburthe1} via a basic uniform model. In the detailed model here, we find that the cooling is also effective, and notably that the process differs significantly from the chromium case. The initial thermal state is prepared in the $m_F = 0$ component, which has lowest quadratic Zeeman energy. Transfer of thermal atoms to the $m_F = \pm 1$ components is possible via the spin-asymmetric contact interactions (the last terms proportional to $c_2$ in Eqs. (\ref{Naequ})).

\subsection{Zero-magnetization case}
\label{zeromagnetization}

\begin{figure}[!bht]
\includegraphics[width=0.3\textwidth, angle=-90, center]{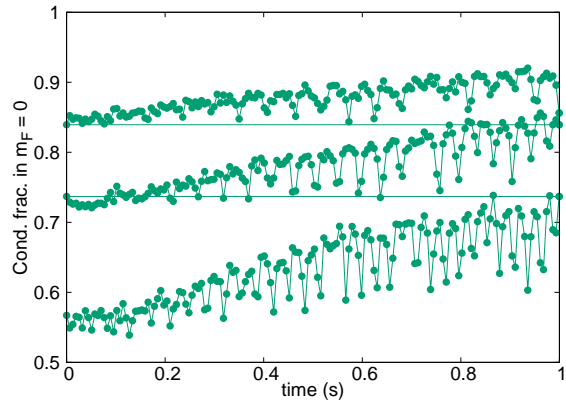}
\caption{Condensate fraction of ${}^{23}$Na in the $m_F = 0$ component as a function of time. Three sets of data represent three consecutive steps of the cooling procedure. The simulation was performed at $q=0.011\hbar\omega_x$ (B$= 100$mG).
}
\label{fig:fig2}
\end{figure}

             We start our analysis with a case of zero magnetization, i.e. when the populations of $m_F=\pm 1$ states are equal. Additionally, we choose initial occupations of these states as almost zero -- only a tiny seed is left to allow a transfer to $m_F=\pm 1$ states. Other starting conditions are possible as well and are mentioned in the next subsection.  The results of a simulation are presented in Fig.~\ref{fig:fig2} for Zeeman energy $q=0.011\hbar \omega_x$, about $N=20000$ atoms and an initial temperature of about $235\,$nK. Significant cooling is seen in each cooling cycle, via an increase of condensate fraction in $m_F = 0$, though notably the timescale is longer than for chromium. Cooling slows after about 1s, which is a good time to remove atoms in the $m_F=\pm1$ states and begin a new cycle. After three cycles the $m_F=0$ population is $N \sim 4500-10000$ (large Rabi oscilations) with 90\% condensate.

\begin{figure}[!bht]
\begin{center}
\includegraphics[width=0.25\textwidth, angle=-90]{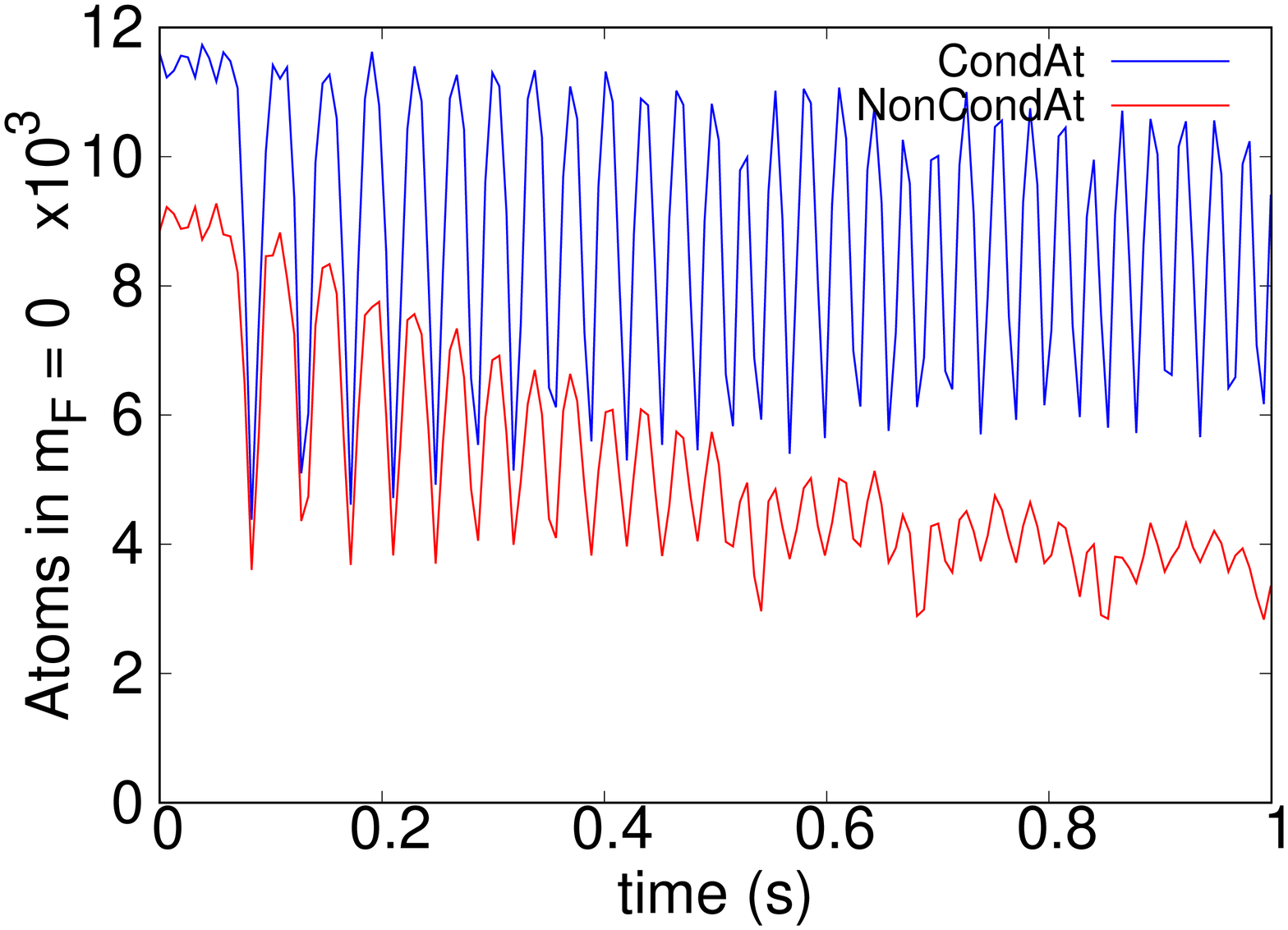}%
\includegraphics[width=0.25\textwidth, angle=-90]{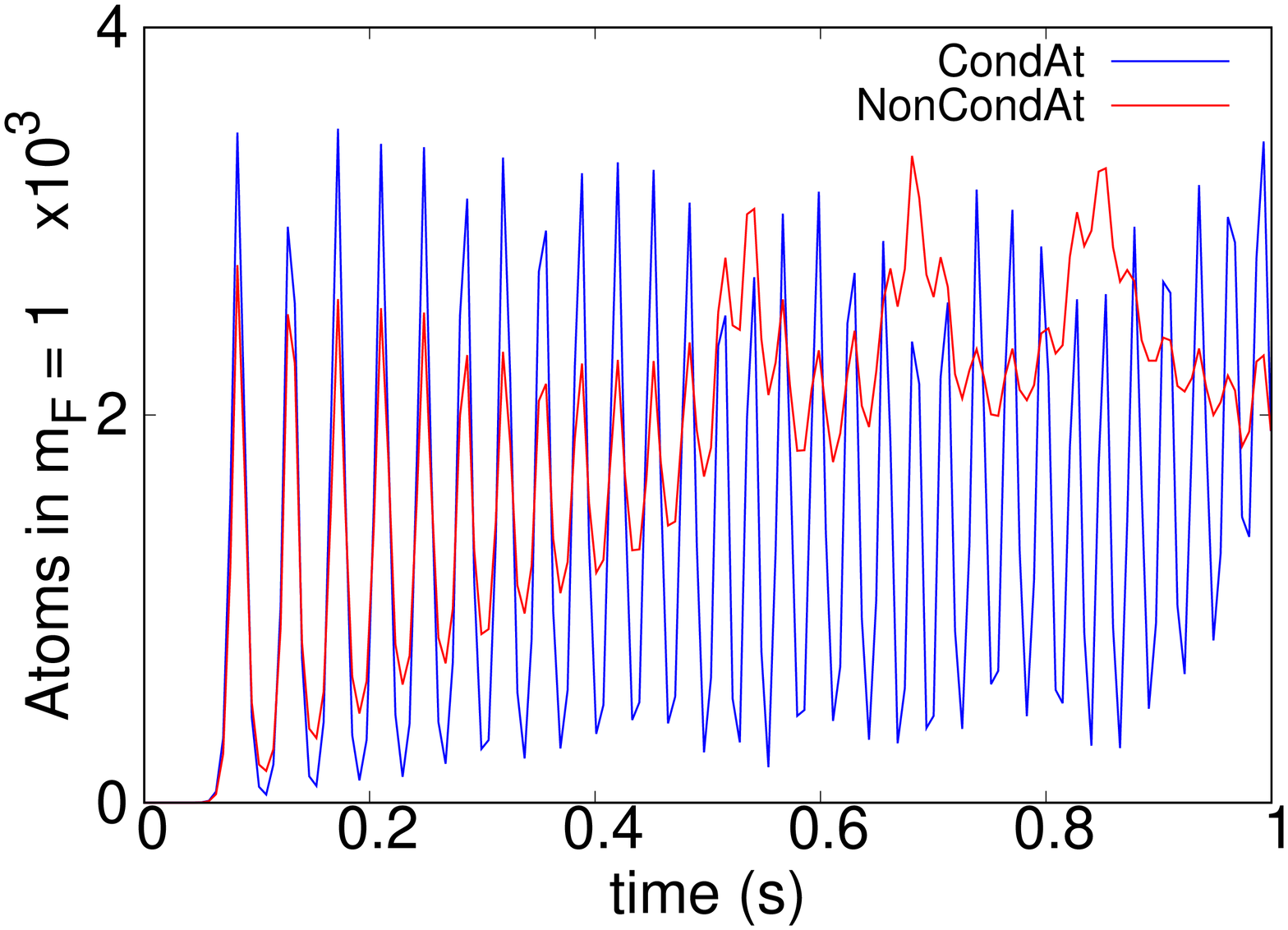}\\
\includegraphics[width=0.25\textwidth, angle=-90]{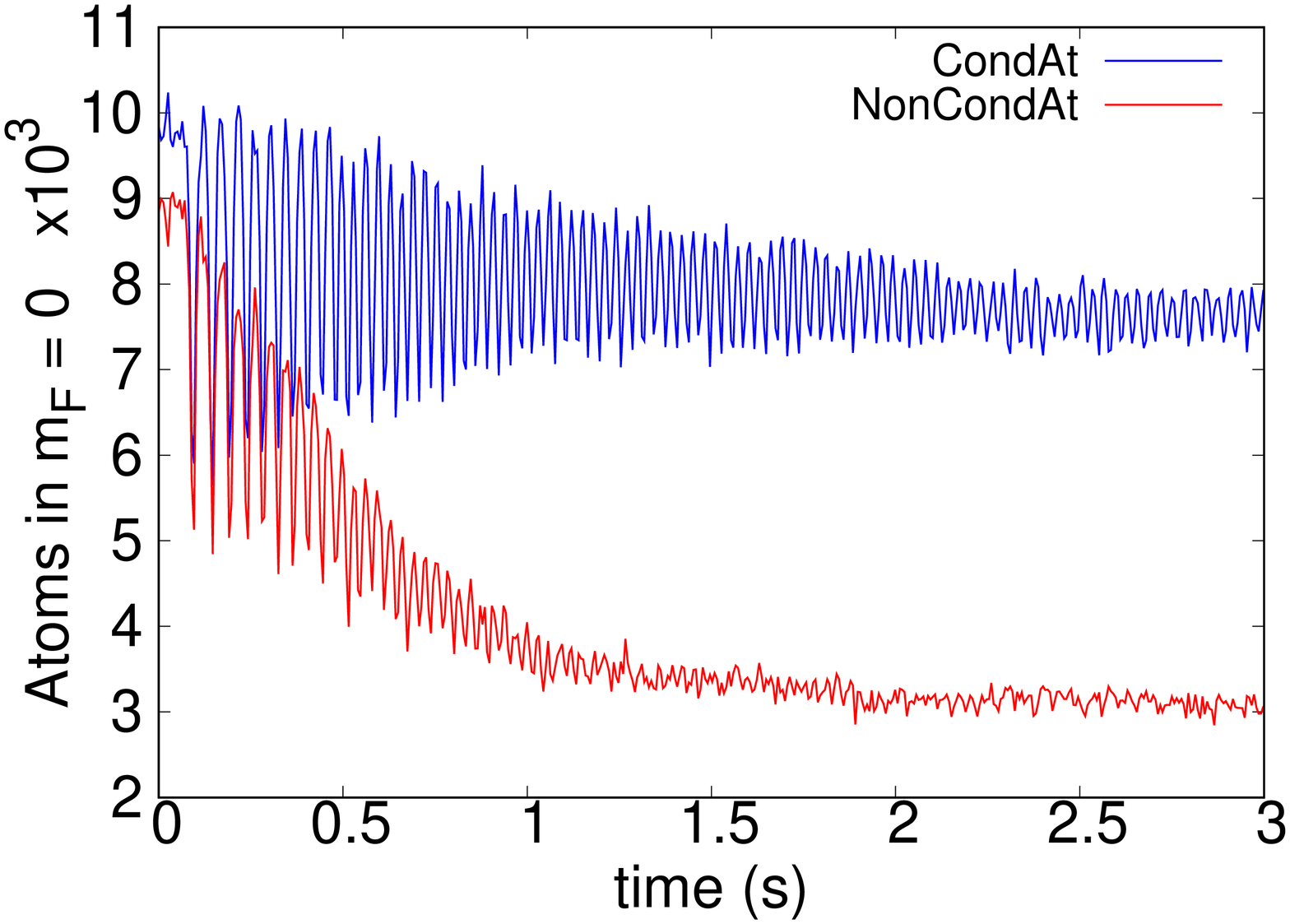}%
\includegraphics[width=0.25\textwidth, angle=-90]{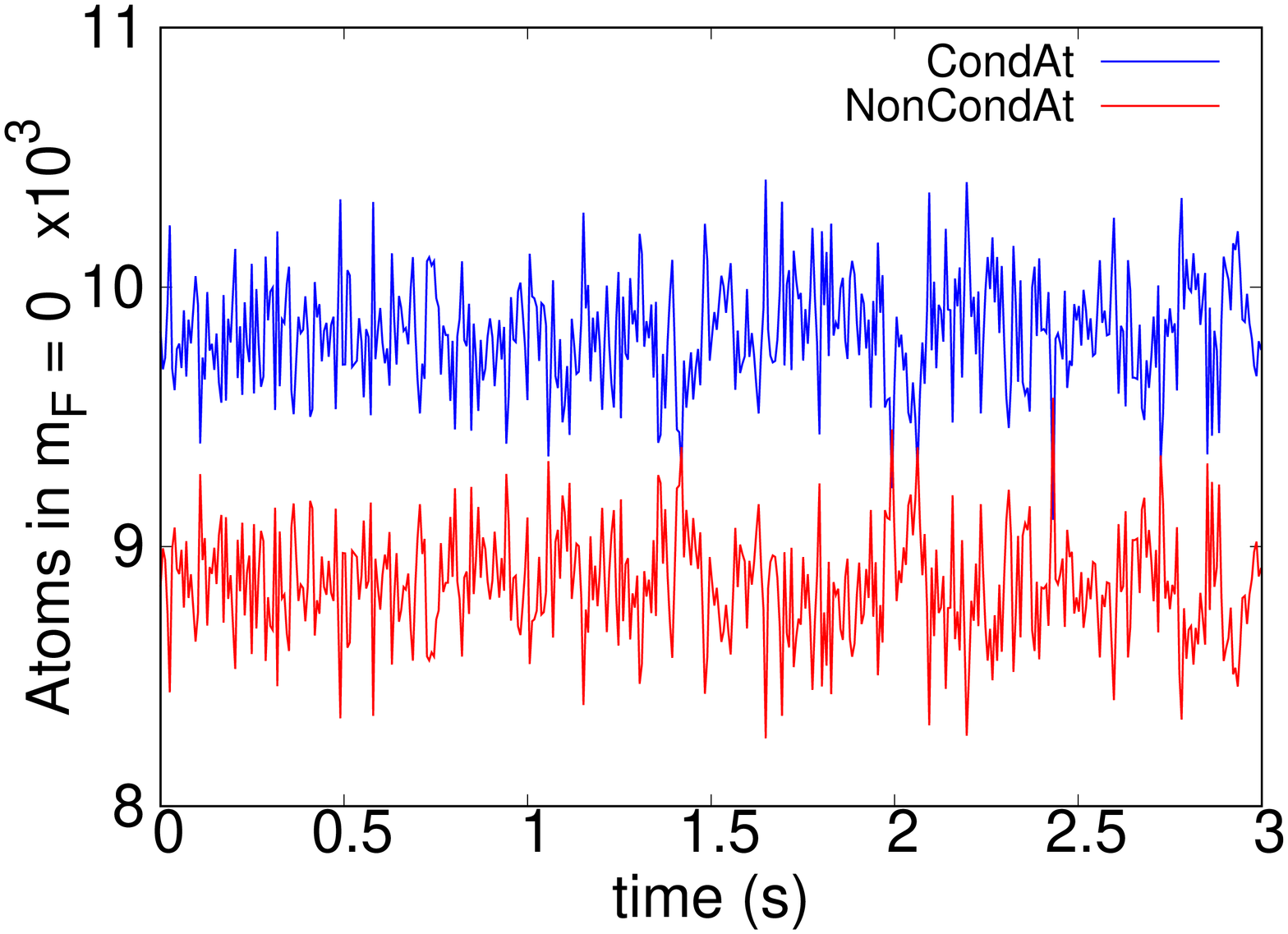}\\
\caption{
Condensate and thermal fraction occupations for ${}^{23}$Na simulations. 
The top row shows the first cycle of the simulation with $q=0.011\hbar\omega_x$ (B=100mG) from Fig.~\ref{fig:fig2}, 
with $m_F=0$ populations on the left, $m_F=+1$ on the right.
The bottom row shows $m_F=0$ populations for: B$= 285$mG, $q=0.09\hbar\omega_x$ on the left (below threshold) and  B$= 425$mG, $q=0.2\hbar\omega_x$ on the right (above threshold).
In all cases condensate occupations are in blue, thermal cloud in red.  
}
\label{fig:Nafig4}
\end{center}
\end{figure}

\begin{figure}[!bht]
\begin{center}
\includegraphics[width=0.3\textwidth, angle=-90]{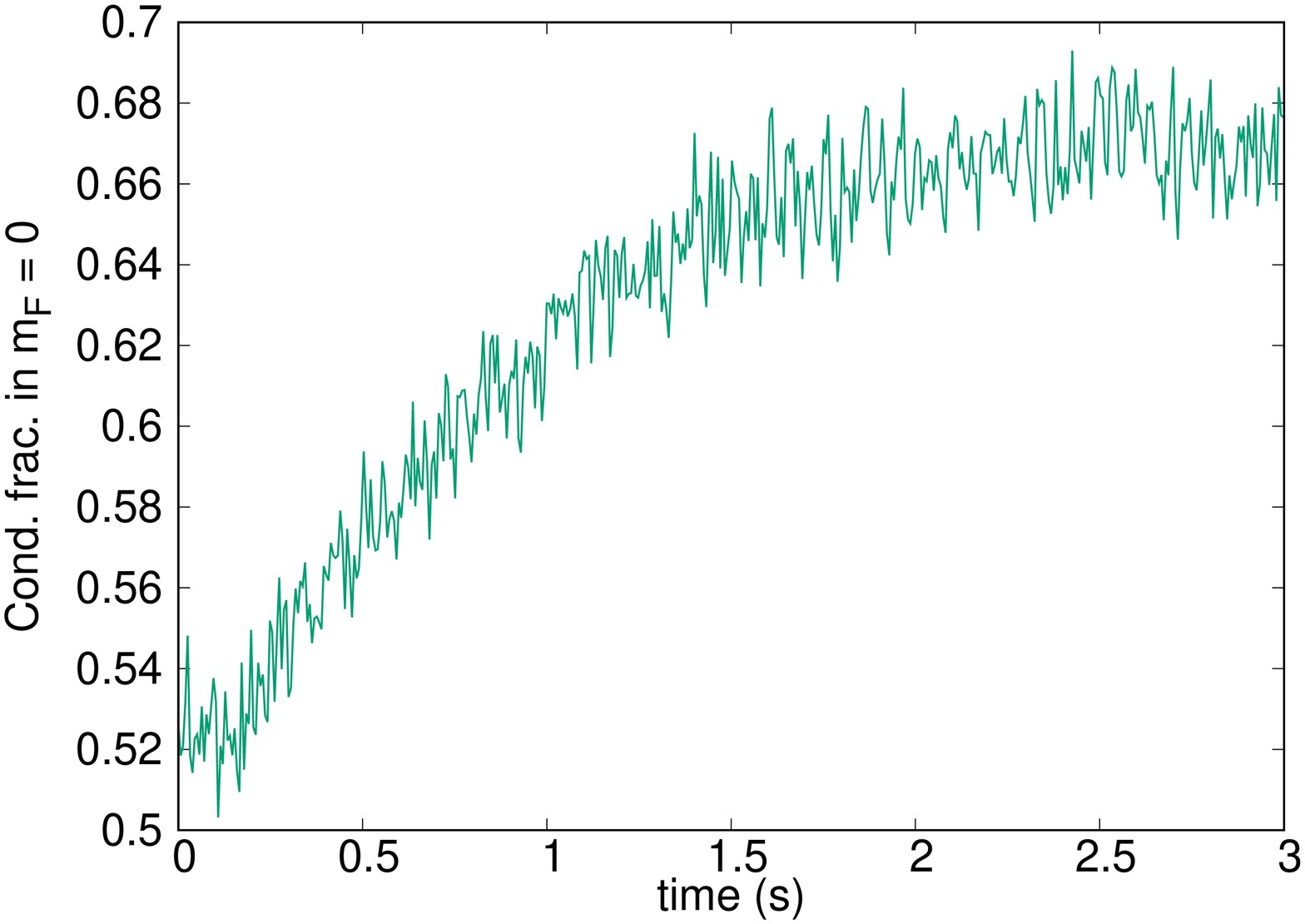}
\includegraphics[width=0.3\textwidth, angle=-90]{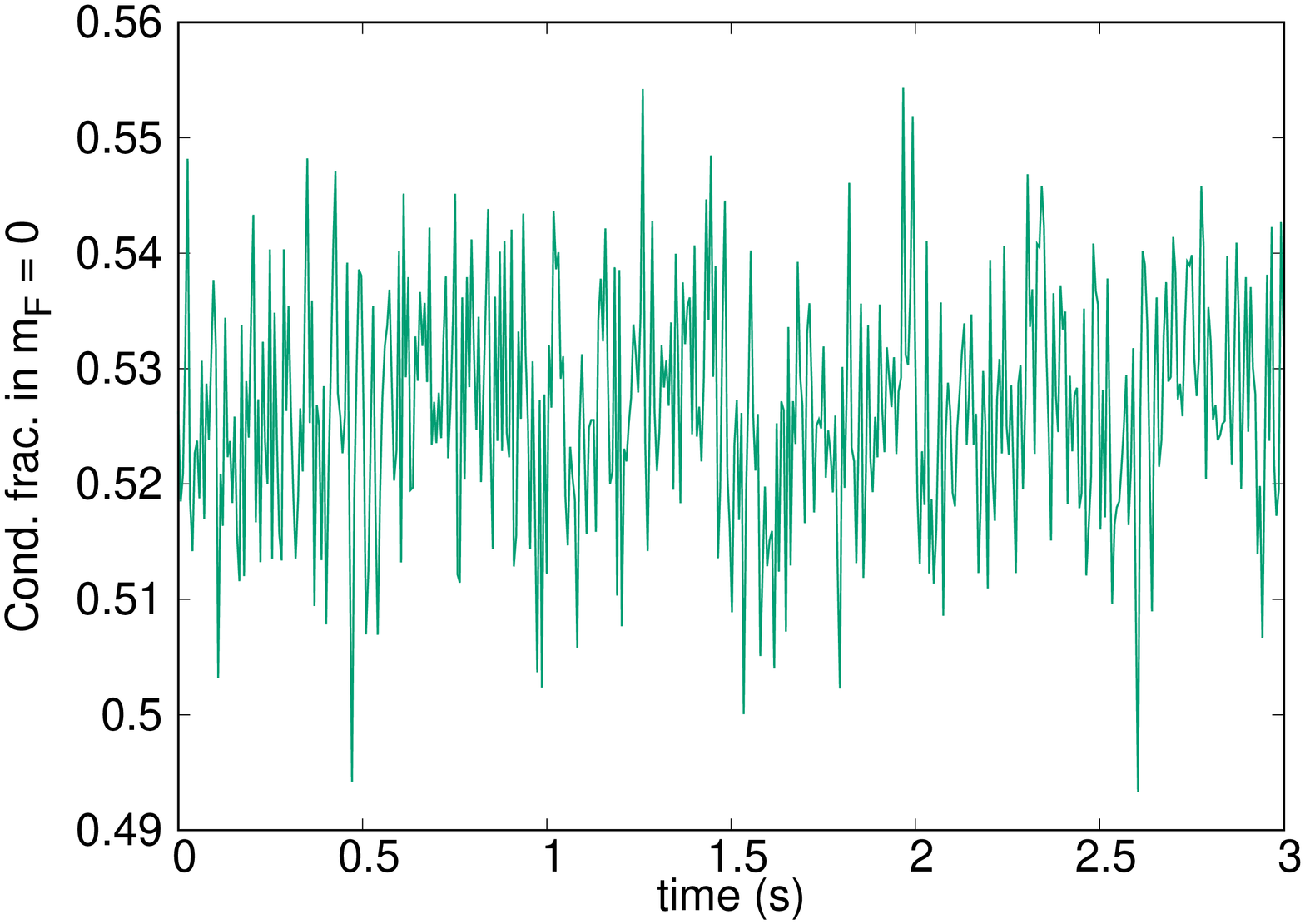}
\caption{Condensate fraction of ${}^{23}$Na in $m_F = 0$ component as a function of time at magnetic fields of B$= 368$mG (left, $q=0.15\hbar\omega_x$) and B$ = 392$mG (right, $q=0.17\hbar\omega_x$). 
}
\label{fig:fig8}
\end{center}
\end{figure}

A visible element of the evolution is a rapid oscillation (period $\sim30$ms). This stems from a Rabi oscillation that coherently transfers both condensate and noncondensate atoms between the $m_F=0$ and $m_F=\pm1$ hyperfine states as can be deduced from the top row of Fig.~\ref{fig:Nafig4}. Its period ($\sim30$ms) corresponds well to predictions based on the energy of the spin changing term $c_2n_0$. The presence of the oscillation indicates coherent exchange between spin states, both in the condensate and the thermal cloud (at least initially). Therefore a different process is dominant than the one seen for chromium.

\begin{figure}[!bht]
\includegraphics[width=0.35\textwidth, angle=-90, center]{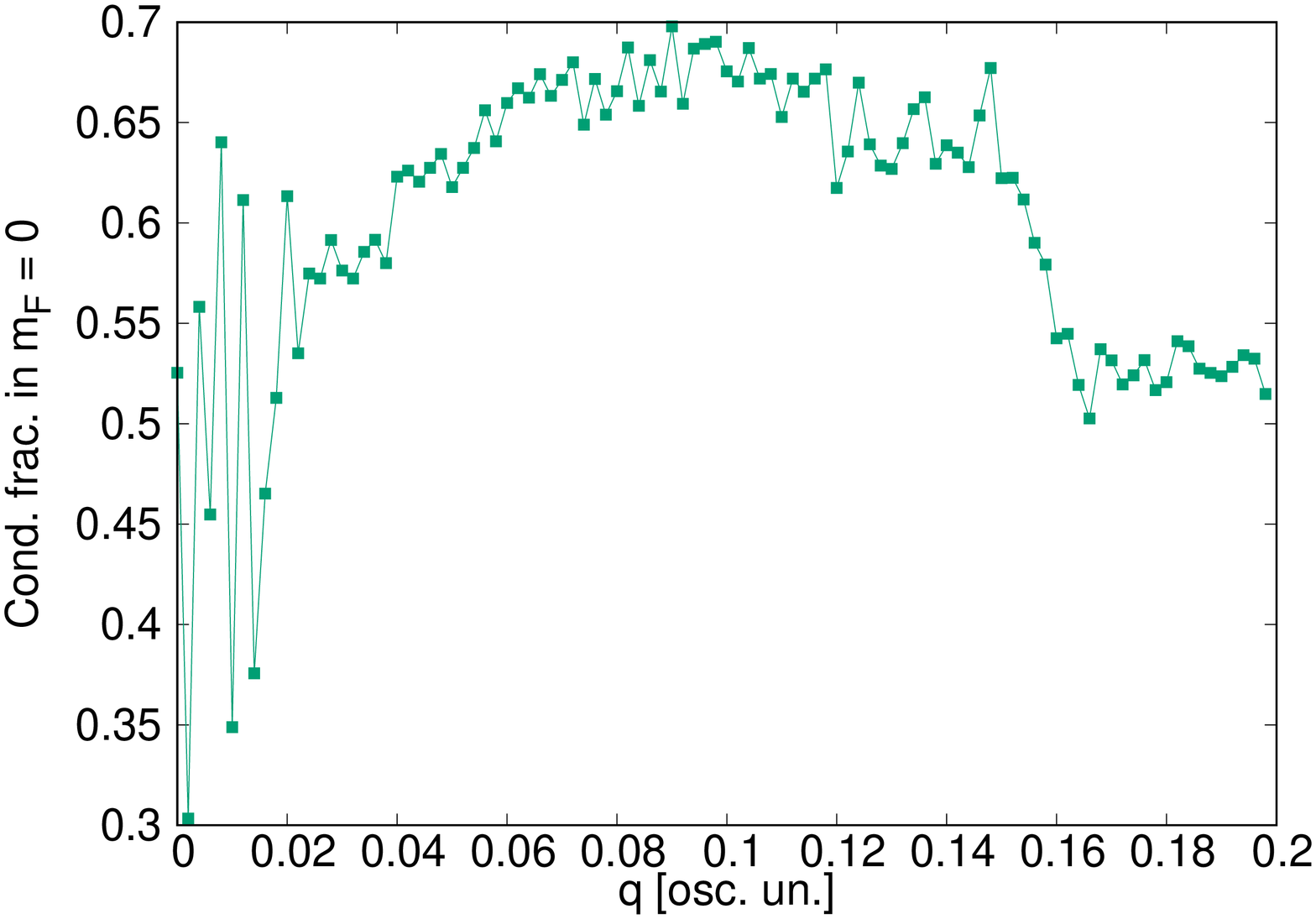}
\caption{
Final ${}^{23}$Na condensate fraction in $m_F=0$ after one              1s cycle,  as a function of the applied quadratic Zeeman energy $q/\hbar\omega_x$. Cooling ceases suddenly above $q=0.15\hbar\omega_x$.
             At the lowest $q\lesssim 0.025$, the 1s experimentally realistic cycle is too short for stabilisation and the randomness reflects snapshots of different stages of the system's Rabi cycle at 1s. An example is seen in Fig.~\ref{fig:Nafig4}, top left. }
\label{fig:vsq}
\end{figure}

The obvious condition to transfer thermal atoms from $m_F = 0$ to $\pm 1$ via the $2 n_0 \rightleftharpoons n_{\pm 1}$ mechanism is that the thermal energy is sufficient to overcome the quadratic Zeeman energy, $q$,  i.e.  $q\lesssim k_BT$. However, the magnetic field used here is already practically negligible compared the thermal energy ($k_BT\sim 1800q$ in Fig.~\ref{fig:fig2}). Instead, we find a limiting factor at much lower magnetic field. Fig. \ref{fig:fig8} shows the first step of cooling for $q/\hbar \omega_x=0.15$ and $0.17$. Surprisingly, no cooling is observed for the higher magnetic field, even though the thermal energy there is still enormously greater than the Zeeman energy: $k_B T/q \approx 100$. Fig.~\ref{fig:vsq} shows the abrupt breakdown of cooling with growing Zeeman energy $q$ in detail. There must be another limitation involved. 

\begin{figure}[!bht]
\begin{center}
\includegraphics[width=0.25\textwidth, angle=-90]{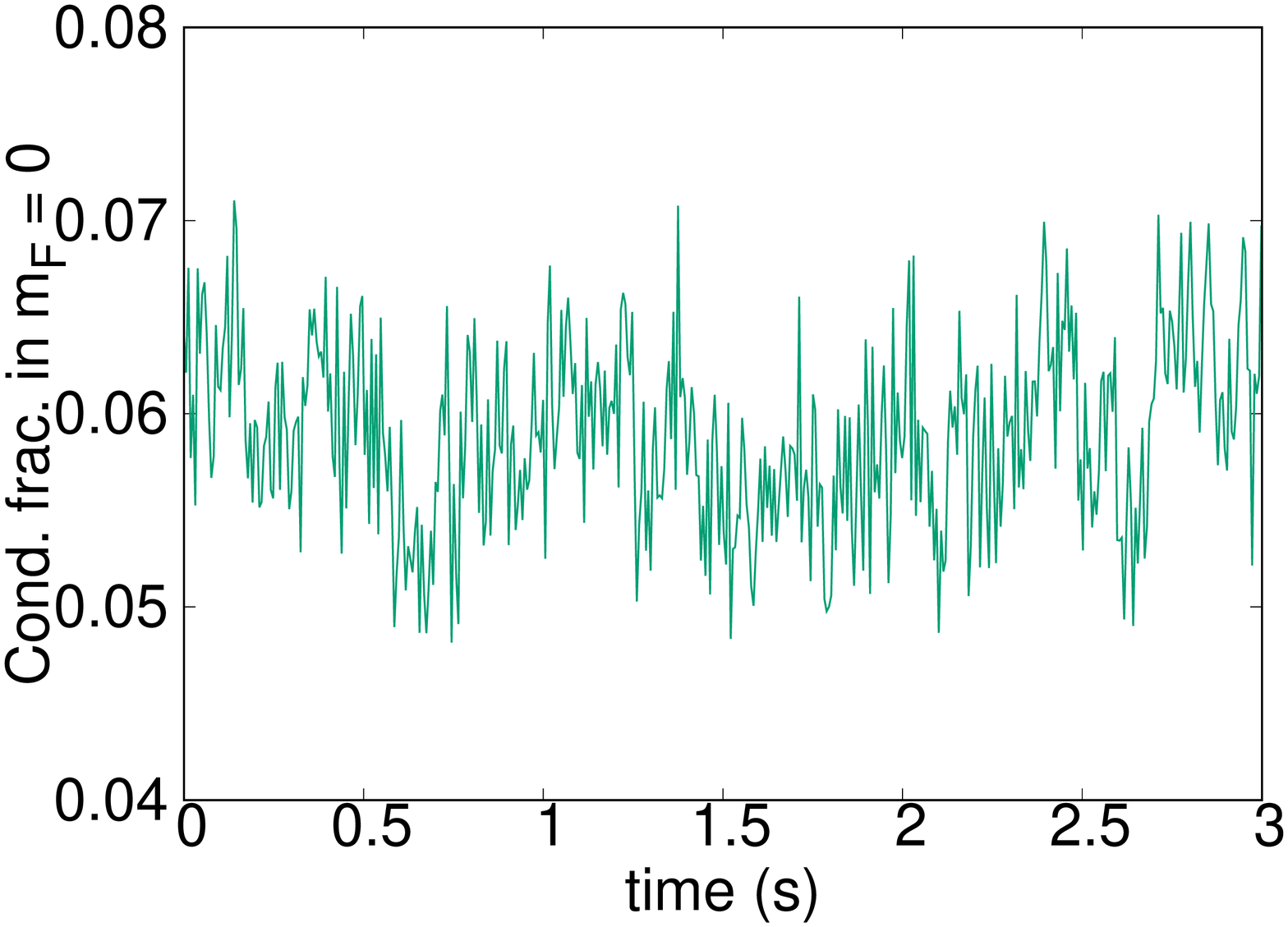}%
\includegraphics[width=0.25\textwidth, angle=-90]{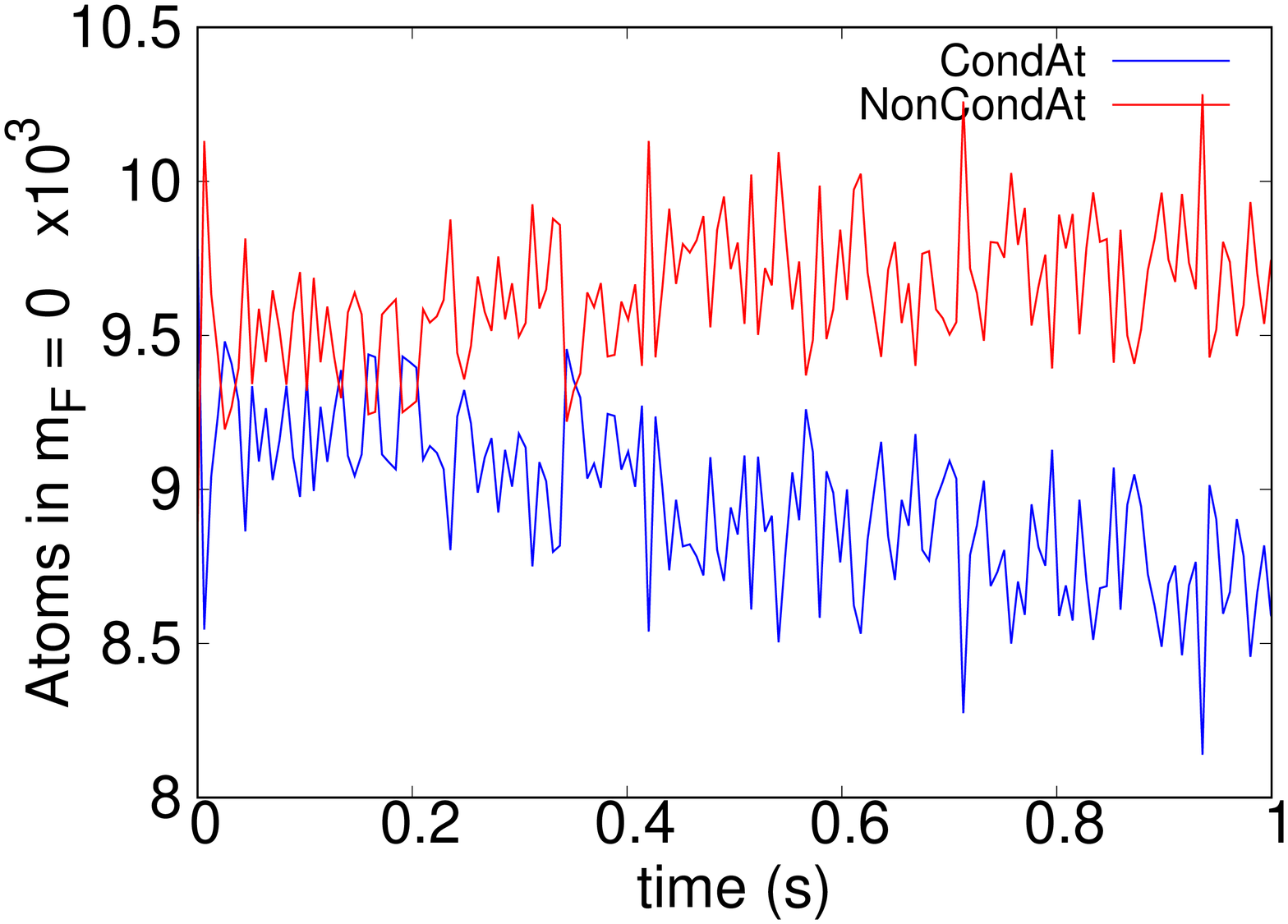}\\
\includegraphics[width=0.25\textwidth, angle=-90]{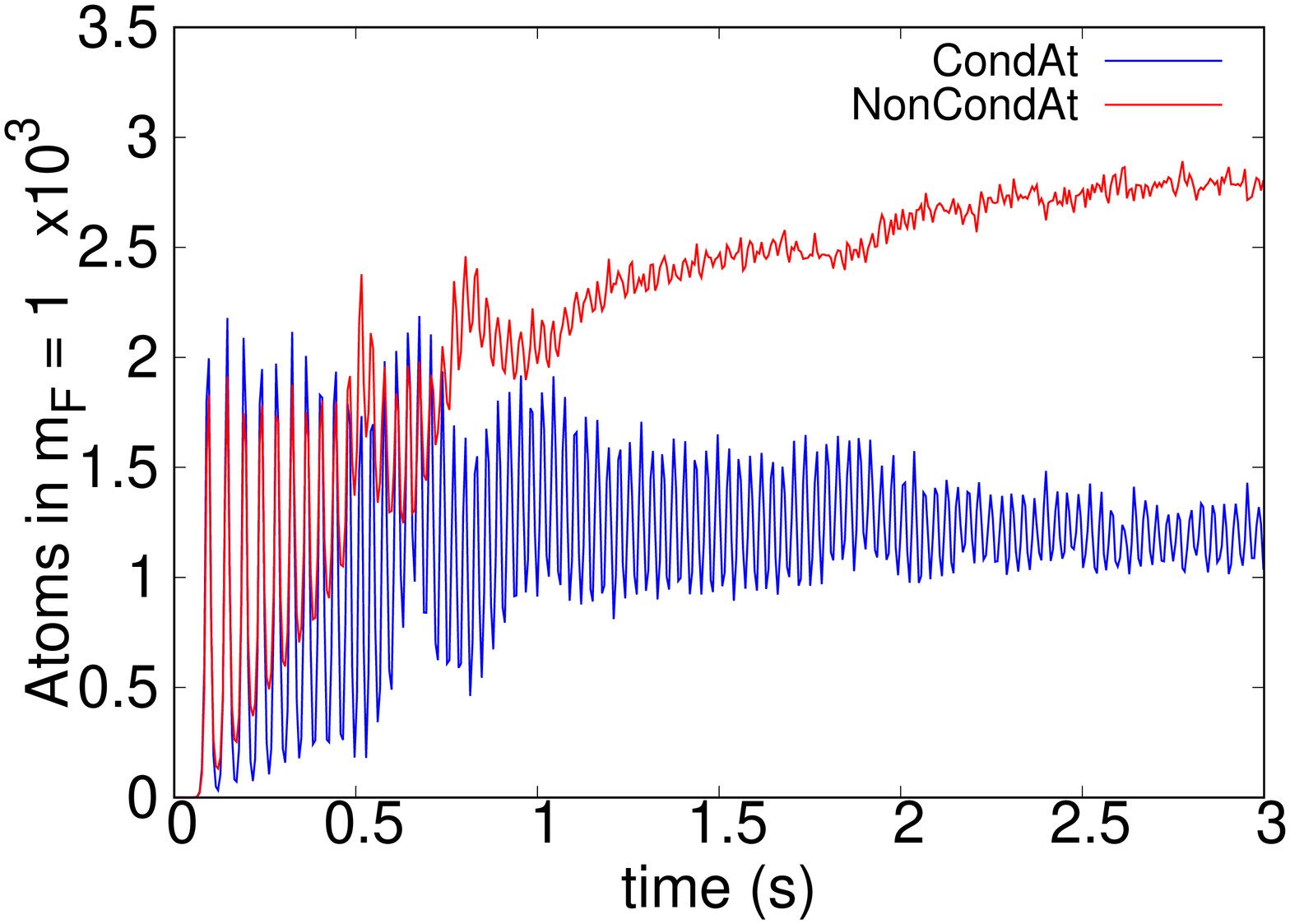}%
\includegraphics[width=0.25\textwidth, angle=-90]{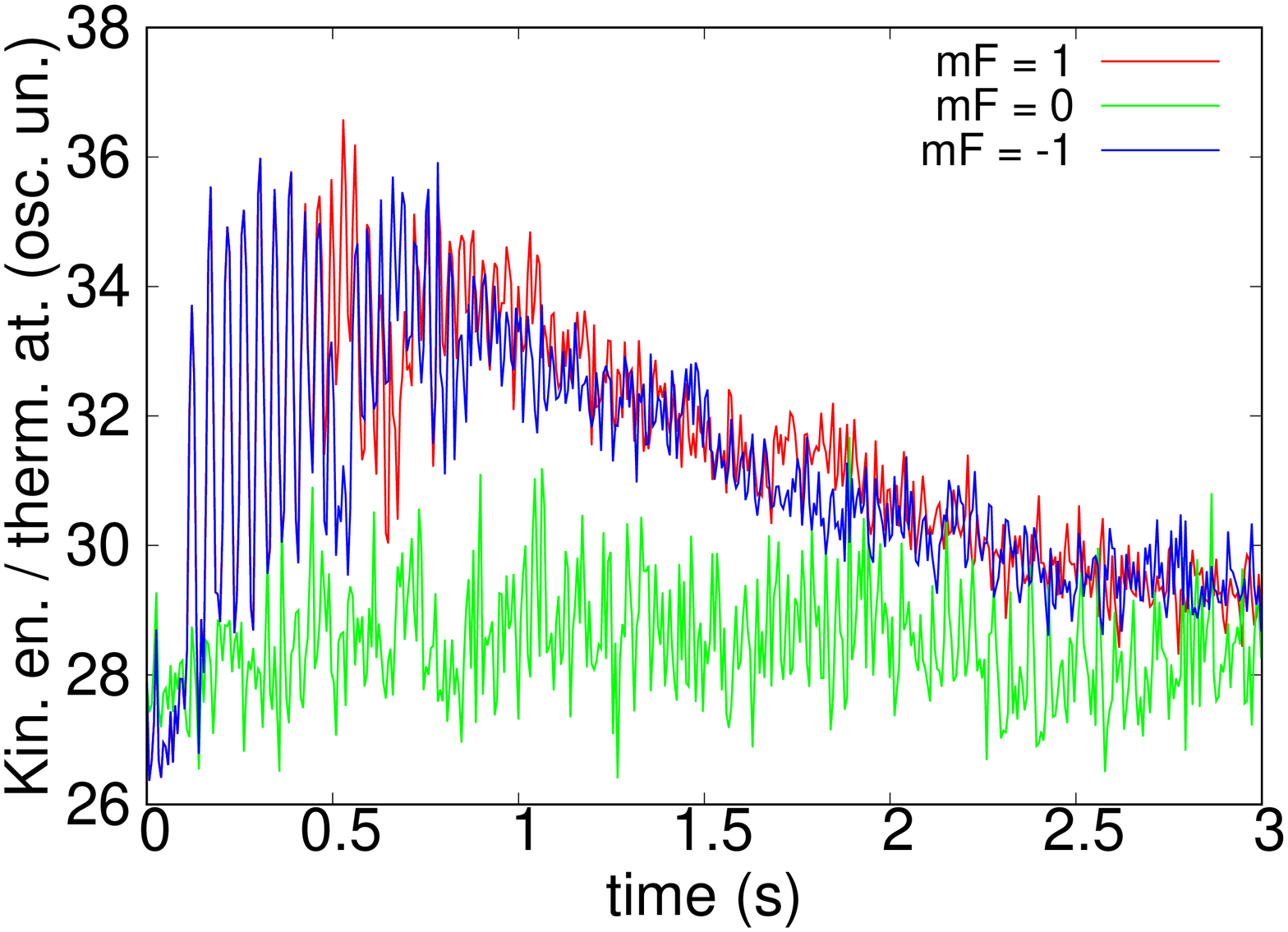}%
\caption{
Some investigations of sodium cooling.  
Top left: Condensate fraction in $m_F=0$ when starting from very low initial values (6\%), $q=0.09\hbar\omega_x$ (B$= 285$mG).
Top right: condensate(blue) / thermal (red) populations in $m_F=0$ when terms responsible for elastic $\psi_0\,\&\psi_{\pm1} \rightleftharpoons \psi_0\,\&\psi_{\pm1}$ collisions were removed ($q=0.011\hbar\omega_x$, B$= 100$mG).
Bottom left: condensate/thermal populations in $m_F=1$ for the $q=0.09\hbar\omega_x$, 55\% initial codensate case shown in Fig.~\ref{fig:Nafig4} (bottom left). 
Bottom right: Kinetic energy per thermal atom in the same case, where red, green and blue lines correspond to the $m_F = 1,0,-1$ components, respectively.
}
\label{fig:4bits}
\end{center}
\end{figure}

The spin mixing terms responsible for transfer from the initial $m_F=0$ to $m_F=\pm1$ have energy of the order of $c_2n_0$. Therefore, the amplitude of any spin mixing process will decay rapidly once the energy difference exceeds this level. In order to take two atoms from $m_F=0$ and place them in the $m_F=\pm1$ states, an energy of {$q$ per atom} is required, leading to the additional condition 
\begin{equation}\label{qc2}
q \lesssim c_2 n_0^{\rm max}
\end{equation}
for cooling onset.
Thresholds of the same order ($q\lesssim\mathcal{O}(c_2n)$) are commonly seen in studies of the stationary states of low temperature  $F=1$ anti-ferromagnetic condensates \cite{Stenger98,Matuszewski,Matuszewski10}. In our case, taking the peak density of a 100\% condensate, condition (\ref{qc2}) is $q \lesssim 0.16\hbar \omega_x$, which matches quite well with the threshold seen in Fig.~\ref{fig:vsq}.
Further corroborating evidence for this interpretation of the cooling threshold includes the fact that no Rabi oscillations are seen above threshold at $q=0.2\hbar\omega_x$ in Fig.~\ref{fig:Nafig4} (bottom right). 
Moreover, if one takes a small condensed fraction of 6\%, no cooling or Rabi oscillations are observed (Fig.~\ref{fig:4bits}, top left), nor transfer to $m_F=\pm1$, despite the same Zeeman energy $q=0.09\hbar\omega_x$ that cooled the larger condensate in Fig.~\ref{fig:fig8}. 
This is also consistent and expected  since a small condensate has highly reduced density, and the condition  (\ref{qc2}) now becomes 
 $q \lesssim 0.03\hbar \omega_x$ -- which is not met.

What is the actual mechanism for cooling, though? It cannot be a threshold Zeeman energy that sorts atoms into high and low energy in different spin states
like we saw in chromium, since the 
atoms can be seen flowing freely to all spin states and back.
On the other hand, the population transfer process is not reversible in the long term and the populations and condensate fractions settle to a more or less stationary value. 
There is another collision process that does not change spin populations, but can rearrange atoms between modes. The spin-preserving and spin-exchanging collisions between $m_F=0$ and $m_F=\pm1$ atoms include contributions such as 
$\psi_0^c\,\&\,\psi_{\pm1}^{\rm th} \leftrightarrow  \psi_0^{\rm th}\,\&\,\psi_{\pm1}^c$ which can exchange thermal and condensate fractions within a spin component, and degrade the Rabi oscillations, leading to a stationary equilibrated state. Notably, the cooling is arrested when the $i\dot{\psi}_0\sim(n_{+1}+n_{-1})\psi_0$ and $i\dot{\psi}_{\pm1}\sim n_0\psi_{\pm1}$ terms responsible for the $\psi_0\,\&\psi_{\pm1} \rightleftharpoons \psi_0\,\&\psi_{\pm1}$ collisions are removed (Fig.~\ref{fig:4bits} top right), proving the crucial role of this process.

The process can be followed in Fig.~\ref{fig:4bits} (bottom).
Two long timescales are evident: Most of the flow of thermal atoms from $m_F=0$ to $m_F=\pm1$ is completed by about 1s, and it is also at this time that the Rabi oscillations of the thermal components die out (also Fig.~\ref{fig:Nafig4}, bottom left).
The Rabi oscillations between condensates last longer, and there is a second timescale of about 3s over which the last thermal atoms accumulate in $m_F=\pm1$ and
the kinetic energy of thermal atoms in all three hyperfine states equilibrates. 

The cooling procedure can in fact be understood as a re-equilibration punctuated by periodic removal of the atoms in the unwanted spin states. 
Once condition (\ref{qc2}) is met, 
the number of accessible thermal degrees of freedom increases threefold. The thermal atoms redistribute, leaving only $\tfrac{1}{3}$ of the original number in $m_F=0$. 
Figs.~\ref{fig:Nafig4} and~\ref{fig:4bits} (bottom left) confirm that the number of condensate atoms in $m_F=0$ reduces only slightly by about 2000, while the initial 9000 thermal atoms in $m_F=0$ redistribute approximately equally among the three hyperfine states, leaving only 3000 in $m_F=0$. Hence, after $m_F=\pm1$ populations are removed, significant cooling has occurred.

Whether re-distribution can occur depends 
on whether condition (\ref{qc2}) is fulfilled. For homogeneous stationary states it can be shown \cite{Matuszewski} that for zero magnetization miscible solutions exist only provided $q < 2 c_2 n$, which resembles (\ref{qc2}). Only then can the spin-changing processes $\psi_0\,\&\psi_0 \rightleftharpoons \psi_1\,\&\psi_{-1}$ for condensed atoms start, so (\ref{qc2}) works as a kind of ignition condition. After the appearance
of condensed atoms in $\pm 1$ components, the real re-equilibration can ensue. Within this stage, the thermal atoms are redistributed to $\pm 1$ states via spin-preserving and spin-exchanging collisions $\psi_0\,\&\psi_{\pm1} \rightleftharpoons \psi_0\,\&\psi_{\pm1}$, and become ready for removal.

The present study is complementary to the work on sodium in \cite{Laburthe1} (supplement) which used a Bogoliubov approximation to investigate the high condensate fraction regime. The present work considers low initial condensate fraction, and we find that the Bogoliubov regime of condensate fraction $\gtrsim 0.9$ can be reached after just several cooling cycles. The two works mesh then at intermediate temperatures, covering the whole range of cooling from near $T_c$ to very low $T$. Cooling cycle efficiency becomes very high in the phonon regime when $T\ll\mu /k_B$ \cite{Laburthe1}, due to a softer phonon spectrum for the $m_F=\pm1$ states ($E_{\mathbf{k}}^{\pm}\approx |\mathbf{k}|\sqrt{c_2n}$) than the $m_F=0$ states ($E_{\mathbf{k}}^{0}\approx |\mathbf{k}|\sqrt{c_0n}$). For a density $n$. Then, below a given energy $E_{\rm max}$, which corresponds to wavenumbers $k^{0,\pm}_{\rm max}=E_{\rm max}/\sqrt{c_{0,2}n}$, respectively, the number of states in volume $V$ is $\mathcal{N}_{0,\pm}\sim (4\pi/3)(V/(2\pi)^3)(k^{0,\pm}_{\rm max})^3$. As a result, the number of accessible states at the beginning of each cooling cycle goes from $\mathcal{N}_0$ to $\mathcal{N}_0+2\mathcal{N}_{\pm}$. This is an increase by a factor of about $1+2(c_0/c_2)^{3/2}$,  rather that merely by $3$ as seen here at higher temperatures. It has been argued that this kind of cooling may be ultimately only limited by quantum depletion effects at extremely low entropies and thermal fractions. 

\begin{figure}[!bht]
\begin{center}
\includegraphics[width=0.3\textwidth, angle=-90]{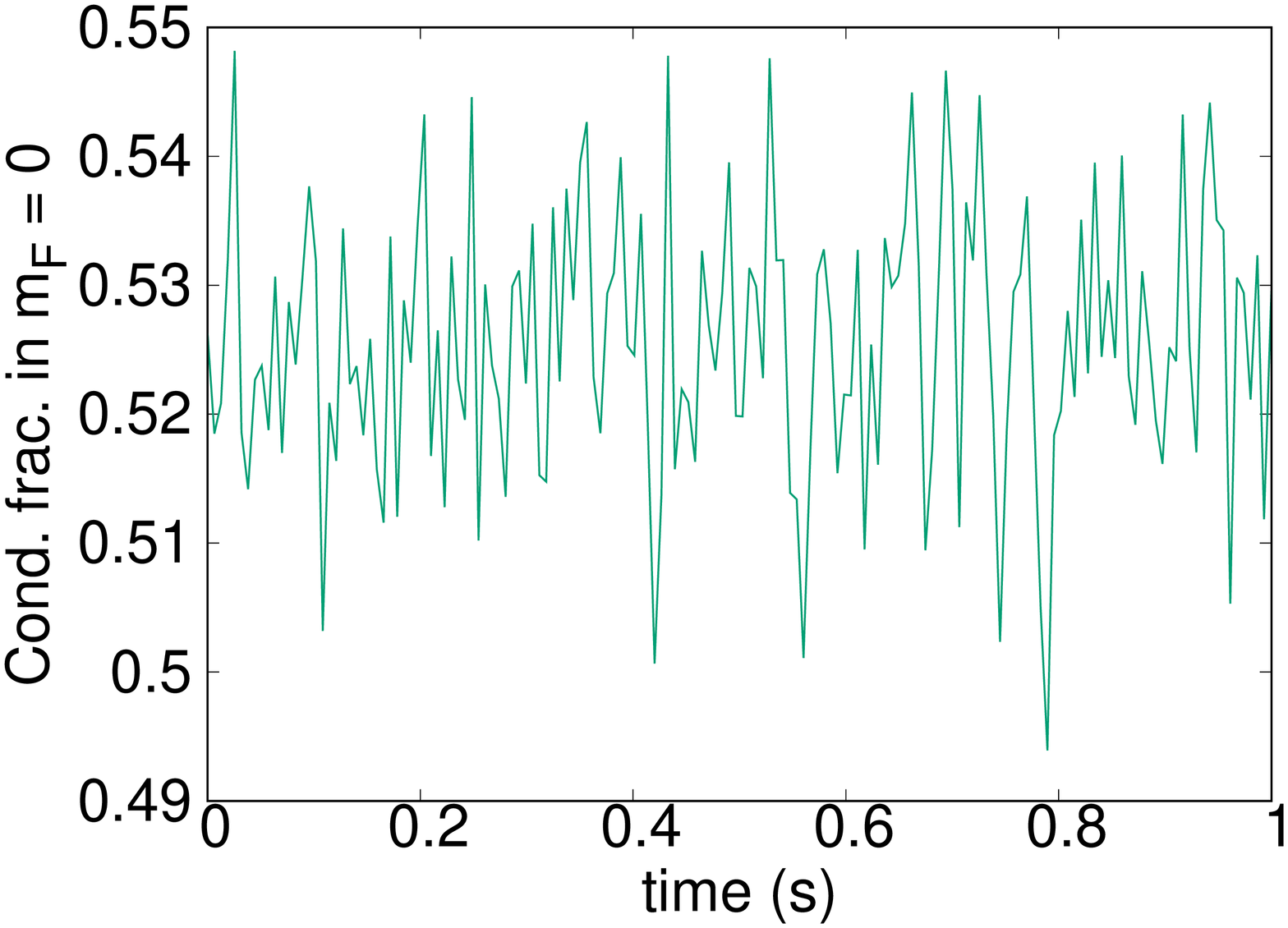}%
\includegraphics[width=0.3\textwidth, angle=-90]{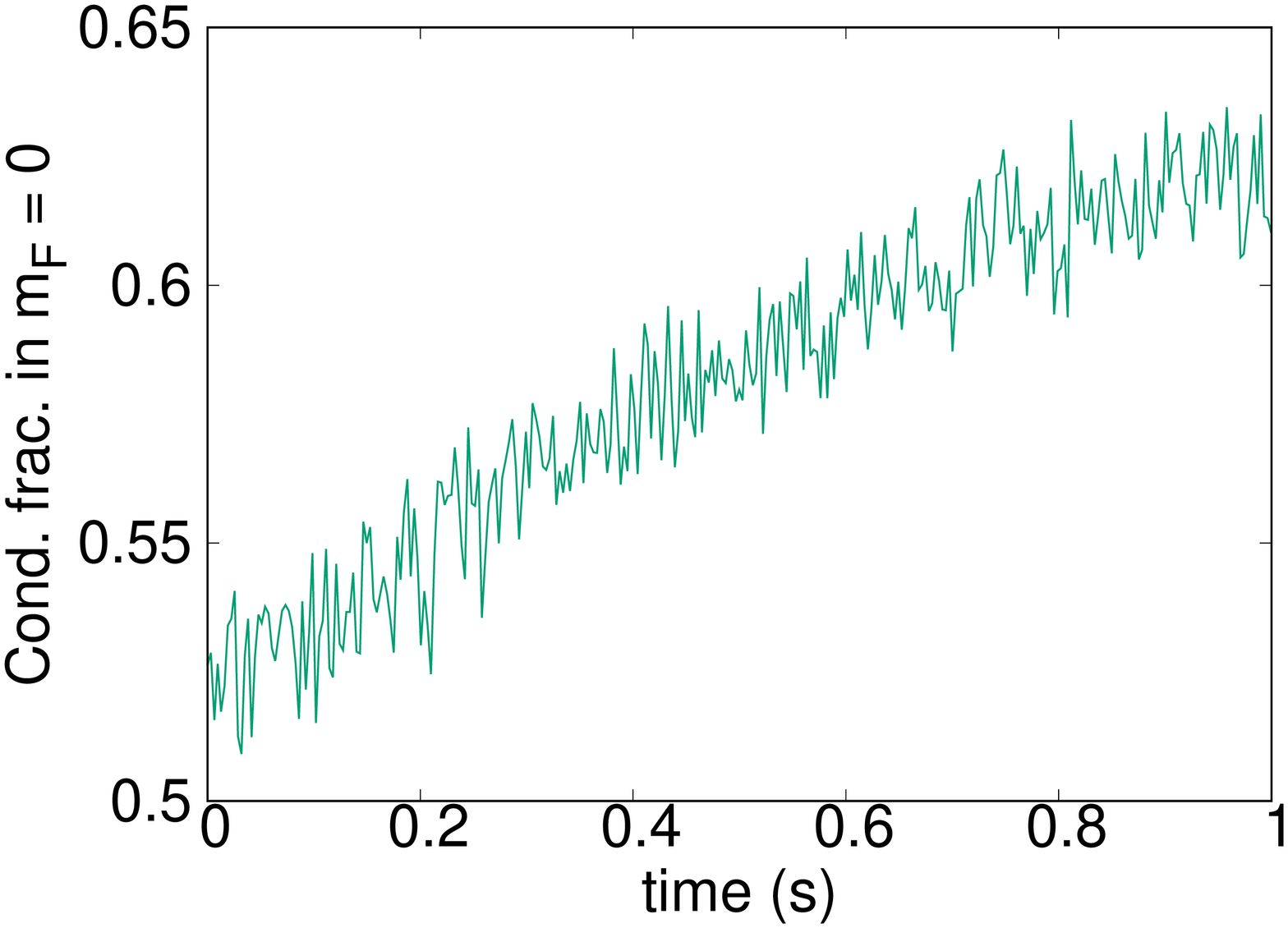}
\caption{Condensate fraction of ${}^{23}$Na in the $m_F = 0$ component as a function of time for a high Zeeman energy of $q=20\hbar\omega_x$ (B$= 4.25G$). This value corresponds to the thermal energy of the system $k_B T$. 
The initial magnetization of the system is $M=0$ (left) and  $M = 0.2$ (right).
}
\label{fig:fig32}
\end{center}
\end{figure}

Finally, we have investigated whether cooling is seen in ${}^{23}$Na  via the process established in Sec.~\ref{Chromiumsection}. The condition (\ref{condCr}) suggests values of $q\approx 20\hbar\omega_x$ to be appropriate as an energy barrier that can only be overcome by thermal atoms. Fig.~\ref{fig:fig32} (left panel) shows the results of such a simulation. Negligible cooling is observed over this timescale. In fact, this is not unexpected,  because there are no dipolar interactions. Therefore, the coherent rapid
process in which a thermal atom in $m_F=0$ is scattered by the condensate to $m_F=1$ or $-1$ cannot take place. 
The incoherent scattering rate can be estimated via $\Gamma=nv\sigma$ with cross-section\cite{Naylor_thesis} $\sigma=4\pi((a_2-a_0)/3)^2$, and thermal relative velocity $v=4\sqrt{k_BT/\pi m}$ \cite{dePaz14}. 
Integratng over space as in \cite{dePaz14} yields a time constant for the $m_F=\pm1$ populations of $\tau=2^{3/2}/\Gamma \sim 15$s, well beyond typical experimental and simulation times.  We expect, though, that a nonzero initial magnetization can help because it opens up other scattering channels.  The right panel of Fig.~\ref{fig:fig32} shows that this is indeed the case.

\subsection{Non-standard initial conditions} 
\label{nonstandard}
To consider values of initial magnetization other than zero, or initial nonzero populations in the $m_F=\pm 1$ components would significantly increase the complexity of the studies, since in such a case the problem becomes dependent actually on three parameters: the quadratic Zeeman energy and the numbers of atoms in $m_F=+1$ and $m_F=-1$ states. On the other hand, as already discussed at the end of previous section and shown in Fig.~\ref{fig:fig32} (right panel) such considerations might open new routes in cooling technique.

\begin{figure}[hbt]
\begin{center}
\includegraphics[width=0.22\textwidth, angle=-90]{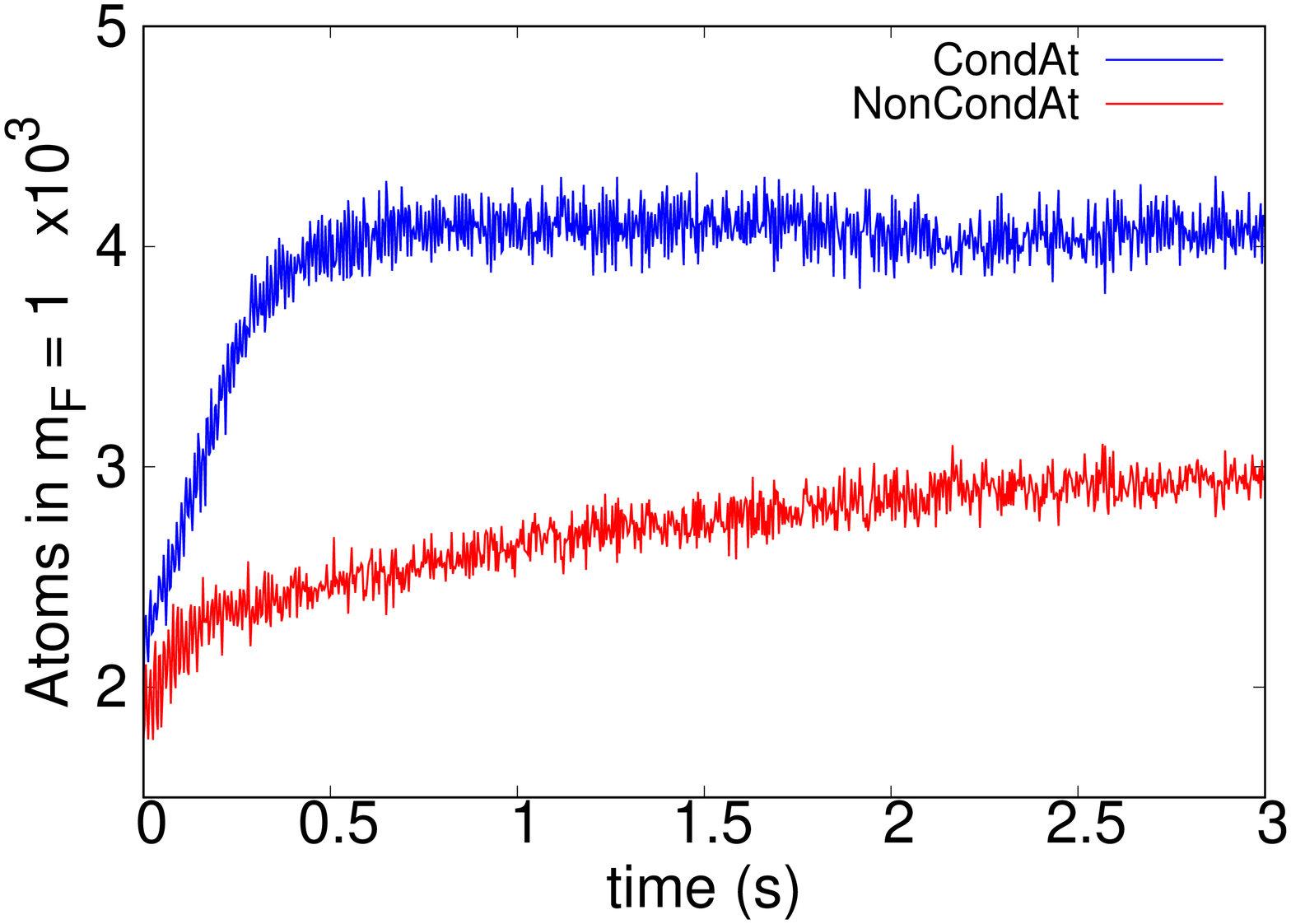}
\includegraphics[width=0.22\textwidth, angle=-90]{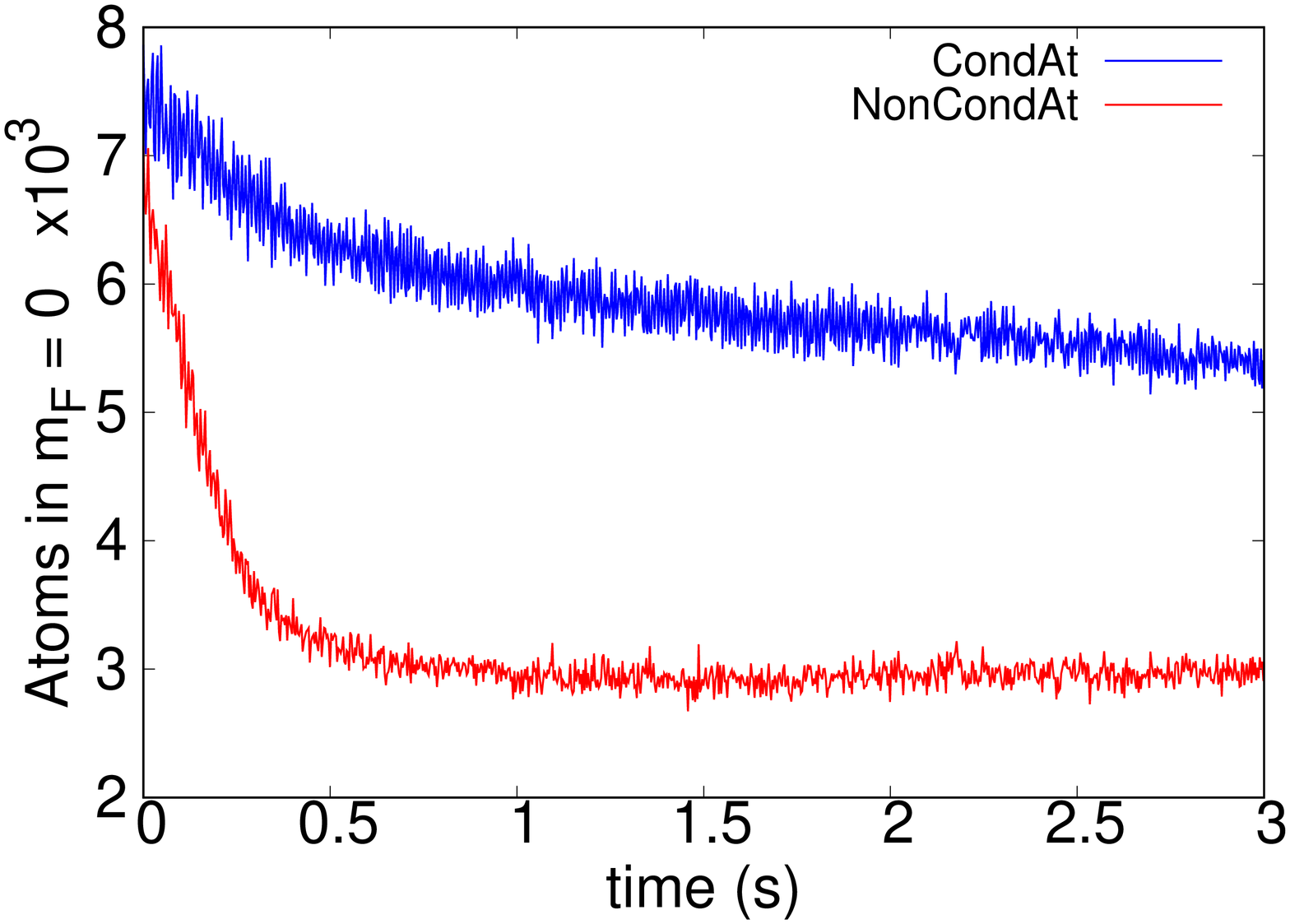}
\includegraphics[width=0.22\textwidth, angle=-90]{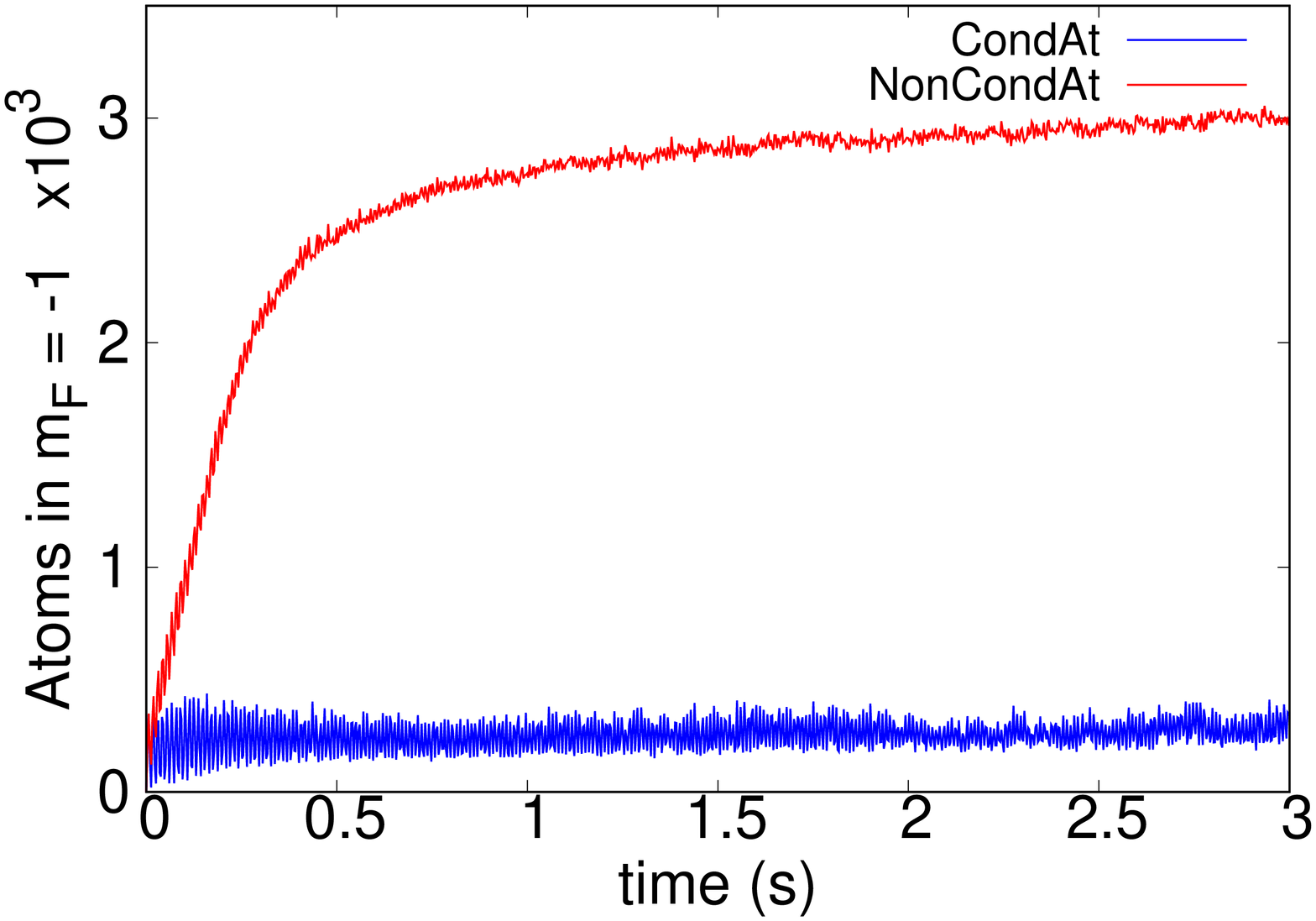}\\
\includegraphics[width=0.22\textwidth, angle=-90]{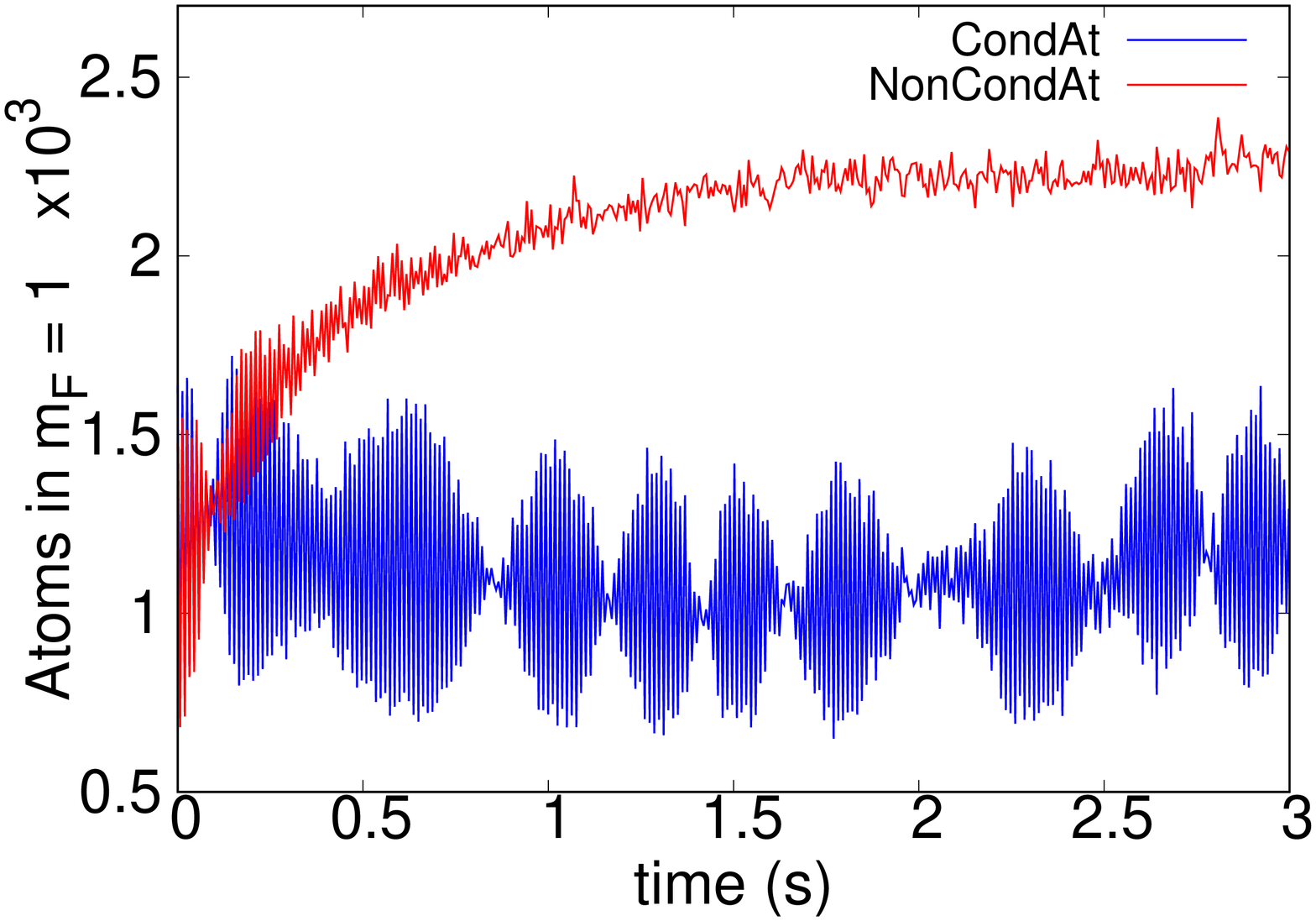}
\includegraphics[width=0.22\textwidth, angle=-90]{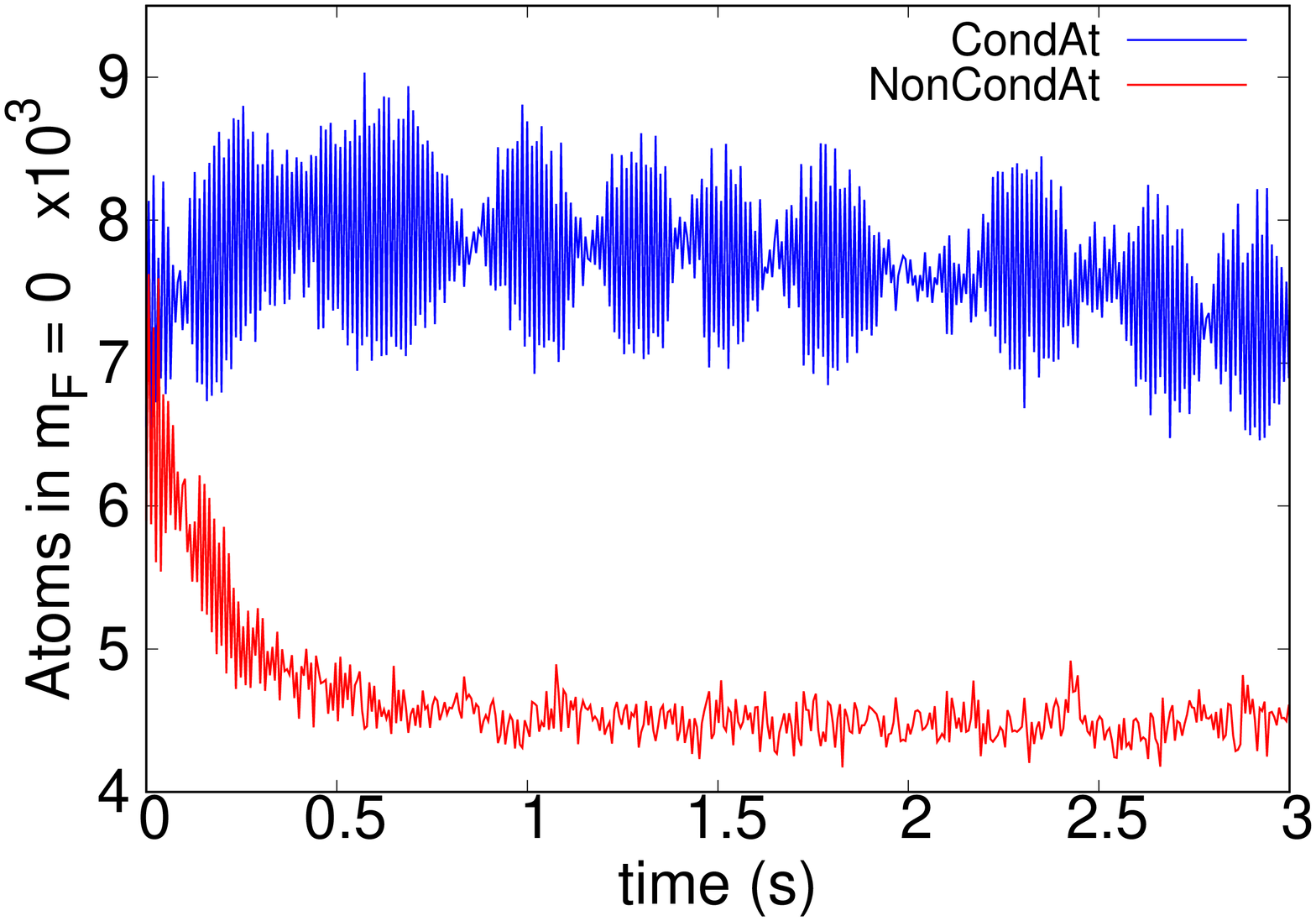}
\includegraphics[width=0.22\textwidth, angle=-90]{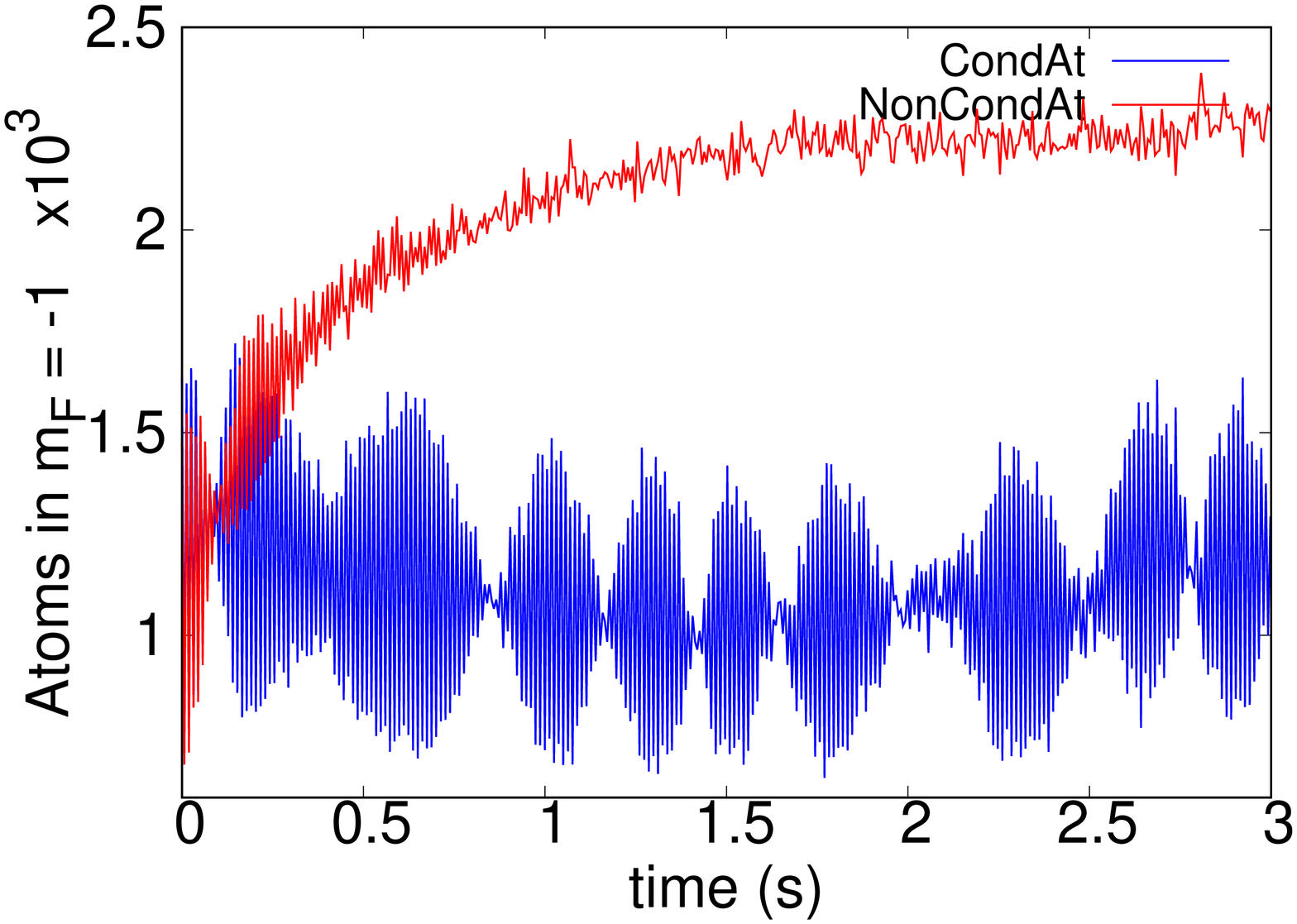}
\caption{
Evolution of condensate and thermal fraction populations for ${}^{23}$Na with $q=0.2\hbar\omega_x$ (B=425mG), but with non-standard starting conditions. 
The top row shows the case with nonzero magnetization (20\% of population initially in $m_F=1$, 80\% in $m_F=0$), while the bottom row the case with $M=0$ but 12\% of initial population in each of $m_F=\pm1$.
In both cases, $m_F=+1$ populations are shown on the left, $m_F=0$ in the middle, $m_F=-1$ on the right. Strong cooling is seen in both cases as evidenced by the drop in non-condensed population in $m_F=0$.
}
\label{fig:fig32-02}
\end{center}
\end{figure}

Leaving detailed analysis of a cooling process in full three-parameter space for a future work, here we just present some intriguing results for the ``non-standard initial conditions'' case.
We have found that if a nonzero magnetization is introduced, or if there is minority occupation of $m_F=\pm1$ states initially, then the range of $q$ that allows cooling increases. Two examples are shown in Fig.~\ref{fig:fig32-02} for $q=0.2\hbar\omega_x$, a value that is too high to lead to cooling when all atoms start in $m_F=0$. Here the initial populations in $m_F=\pm1$ are obtained by coherently transferring part of the initial cloud from the $m_F=0$ hyperfine state at $t=0$, e.g. by applying an appropriate short Bragg pulse to the cloud \cite{Vogels02}. Why cooling appears at higher $q$ here is not obvious and remains an open question. It may, however provide an additional useful pathway. For example, the case with nonzero seed in $m_F=1$ can still be cycled by emptying only the $m_F=-1$ spin state, or cooling at a given $q$ could be started by producing small seeds at $m_F=\pm1$ with a Bragg pulse.

\section{Conclusions}\label{Conclusions}

The spin distillation cooling mechanism as originally proposed and partially experimentally tested in \cite{Laburthe1}, was studied using advanced numerical models that allow for a realistic treatment of temperature, Bose enhancement, and nonlinear dynamics over experimental timescales. 
These calculations confirm that successive cycles of cooling can be applied to a spinor gas, 
 leading to a very high final condensate fraction (97\% 
in our case), while simultaneously losing only a minority of atoms.
Spin distillation was observed both for systems with long-range ($^{52}$Cr, 7 spin states) and short-range ($^{23}$Na, 3 spin states) forces.
We identify two different mechanisms for
a cooling cycle: 

\{1\} 
With a high Zeeman energy barrier, as seen in the chromium example, the thermal atoms dipolarly scattering with condensed ones are able to surmount the energy barrier. Such a mechanism moves thermal energy out of the initial cloud to relatively few atoms in higher spin states,  from which they and the entropy they carry with them can be cyclically removed. This process works for magnetic fields \emph{above} a threshold (\ref{Bth}), and is eventually limited to temperatures of (\ref{limit1}).  Notably, these can be very low temperatures far below the $k_BT\sim\mu$ that is possible with standard evaporative cooling. Also, since the scattered atoms are in a different mode space from the source cloud, removal does not eat into the condensate like in standard evaporation.  

\{2\} 
With a very low Zeeman energy barrier, as seen in the sodium example, an initially spin polarised thermal cloud relaxes into $\tilde{m}$ empty hyperfine states, leaving only $1/(\tilde{m}+1)$ of it in the original spin state. 
Here the cooling can be ignited only below a maximum threshold for the magnetic field (\ref{qc2}), and is not highly sensitive to the value of the temperature. The process can work at very small magnetic fields, far below the bounds set by (\ref{Bth}) and appears quite universal in its simplicity. 

There are similarities in the mechanisms we have identified. In both of them the existence of a condensate is crucial. For chromium, thanks to Bose enhancement, thermal atoms are scattering off condensed ones to higher spin states, state by state. For sodium, coherent transfer of already condensed atoms is fast and triggers the process of redistribution of thermal atoms. Also, in both mechanisms, the role of contact interactions in thermalization of newly populated states is similar. In the chromium case, they allow thermal atoms to reach equilibrium within, but not between, spin components.
For sodium they arrest the Rabi oscillations of condensed and noncondensed atoms between components, first making oscillations of thermal atoms die out and finally decohering condensed and noncondensed atoms.

              Our cooling protocol makes some idealisations. 
For instance, contrarily to what happens in real experiments, we assume perfect and instantaneous removal of unwanted atoms. This idealization can be partially relaxed, simply by trying simulations where a small remnant fraction of atoms is left in $m_s=-2,-1,...,3$ for chromium and $m_F=\pm 1$ components for sodium after each cooling cycle. We have checked that the cooling efficiency remains unchanged to within the noise level in the cooling plots even for a fraction of retained atoms equal to about $10\%$.

The efficiency of evaporative cooling
is commonly characterised by the parameter $\gamma=-d(\ln D_{fin}/D_{ini})/d(\ln N_{fin}/N_{ini})$ \cite{Ketterle96}, where $D_{ini}$ ($D_{fin}$) is the initial (final) phase-space density of a gas and $N_{ini}$ and $N_{fin}$ are the corresponding numbers of atoms. The cooling efficiency in the first cooling cycle can be calculated assuming $D_{fin}$ and $N_{fin}$ are taken at the end of the cycle and concern chromium and sodium atoms in the $m_s=-3$ and $m_F=0$ components, respectively. In our case, to estimate the phase-space density we take the product of the spatial extents of the atomic cloud in the $x$, $y$, and $z$ directions with the respective extents in momentum space available via Fourier transform of the classical field. The classical field is averaged over a short period of time to diminish irrelevant fluctuations before determining the size of the cloud (as the full size at half maximum). 
As a result, we obtain a value of parameter $\gamma \approx 1$ both for chromium and sodium atoms, which is close to typical values of $\gamma$ for alkali atoms for evaporation cooling \cite{Ketterle96}.
 (For example $\gamma=0.86$ for chromium at 1.75 mG and $\gamma=0.53$ for sodium at 100 mG).

In summary, spin distillation cooling emerges as a versatile technique to reach ultra low temperatures in spinor gases, below what is achievable via standard evaporative cooling. It appears to be applicable in a wide range of regimes and for many atomic species, acting through at least two distinct physical mechanisms. The simulations presented above verify that it can be effective under realistic conditions.

\section{Methods}
\label{ddmel}

{\large Matrix elements for the dipolar Hamiltonian}
\newline
The dipolar matrix element $[H_{d}]_{11}$ is given in CGS units by \cite{Swislocki14} 
\begin{align} \label{Hd11}
[H_d (\mathbf{r})]_{11} =\,& {(\hbar\gamma_{\text{\tiny{\rm Cr}}})^2} \int d{\bf r}'  \left[
\frac{1}{|\mathbf{r}-\mathbf{r}'|^3}-3\frac{(z-z')^2}{|\mathbf{r}-\mathbf{r}'|^5}
\right]   \sum_{m=-s}^s  m |\psi_m ({\bf r}')|^2    \nonumber \\
 &-3\, {(\hbar\gamma_{\text{\tiny{\rm Cr}}})^2}
\int d{\bf r}' \frac{z-z'}{|\mathbf{r}-\mathbf{r}'|^5}[(x-x') - i (y-y')]  
\sum_{m=-s+1}^s \sqrt{(4-m)(3+m)/4}\,\, \psi_m^{*} ({\bf r}')\, \psi_{m-1} ({\bf r}')  \nonumber \\
 &-3\, {(\hbar\gamma_{\text{\tiny{\rm Cr}}})^2}
\int d{\bf r}' \frac{z-z'}{|\mathbf{r}-\mathbf{r}'|^5}[(x-x') + i (y-y')] 
\sum_{m=-s}^{s-1} \sqrt{(3-m)(4+m)/4}\,\, \psi_m^{*} ({\bf r}')\, \psi_{m+1} ({\bf r}')   \,,
\end{align}
where $\gamma_{\text{\tiny{\rm Cr}}}=g_L\, \mu_B{/\hbar}$ is the gyromagnetic ratio for ${}^{52}$Cr.
The orthonormal set of spatial coordinates $\mathbf{r}=[x,y,z]$ are arranged such that $z$ lies along the direction of the applied magnetic field ${\bf B}$. 
The off-diagonal dipolar matrix element $[H_{d}]_{10}$ is
\begin{align} \label{Hd10}
[H_d (\mathbf{r})]_{10} =  
&-3 \sqrt{3}\,  {(\hbar\gamma_{\text{\tiny{\rm Cr}}})^2} \int d{\bf r}' \frac{[(x-x')-i(y-y')](z-z')}{|\mathbf{r}-\mathbf{r}'|^5}
\sum_{m=-s}^s  m |\psi_m ({\bf r}')|^2   \nonumber \\
& -3 \sqrt{3}\, {(\hbar\gamma_{\text{\tiny{\rm Cr}}})^2}
\int d{\bf r}' \frac{[(x-x')-i(y-y')]^2}{|\mathbf{r}-\mathbf{r}'|^5}  
\sum_{m=-s+1}^s \sqrt{(4-m)(3+m)/4}\,\, \psi_m^{*} ({\bf r}')\, \psi_{m-1} ({\bf r}')  \nonumber \\
& +\sqrt{3}\,  {(\hbar\gamma_{\text{\tiny{\rm Cr}}})^2} \int d{\bf r}' \left[ \frac{2}{|\mathbf{r}-\mathbf{r}'|^3}-3\frac{(x-x')^2+(y-y')^2}
{|\mathbf{r}-\mathbf{r}'|^5} \right]  
\sum_{m=-s}^{s-1} \sqrt{(3-m)(4+m)/4}\,\, \psi_m^{*} ({\bf r}')\, \psi_{m+1} ({\bf r}')  \;.
\end{align}

\section*{Acknowledgements}
We would like to acknowledge helpful discussions with Emilia Witkowska, and especially Bruno Laburthe-Tolra who pointed out to us a number of important aspects,
 and are grateful to Joanna Pietraszewicz for early related studies. We acknowledge support from the National Science Center (Poland): 
TS from grant No. 2015/17/D/ST2/03527; PD from grant No. 2018/31/B/ST2/01871. MB and MG received funding from project MAQS. Project MAQS is supported by the National Science Centre, Poland under QuantERA, which has received funding from the European Union’s Horizon 2020 research and innovation programme under grant agreement no 731473.

\section*{Author contributions statement}

All authors made essential contributions to the work, discussed results, and contributed to the writing of the manuscript. The numerical simulations were performed
by T.S.



\end{document}